\documentclass[sn-mathphys]{sn-jnl}

\usepackage{array}
\usepackage[para]{threeparttable}
\usepackage{booktabs, caption, makecell}

\jyear{2023}%

\newcommand{\pT}{p^T_t}
\newcommand{\mtt}{M_{t\bar{t}}}
\newcommand{\yt}{y_{t}}
\newcommand{\ytt}{y_{t\bar{t}}}
\newcommand{\mt}{m_t}
\newcommand{\ttbar}{t\bar{t}}
\newcommand{\as}{\alpha_s(M_Z)}
\newcommand{\gev}{\mathrm{GeV}}

\usepackage{eso-pic}
\newcommand{\reportnum}[2]{
  \AddToShipoutPictureBG*{%
    \AtPageUpperLeft{%
      \hspace{0.75\paperwidth}%
      \raisebox{#1\baselineskip}{%
        \makebox[0pt][l]{\textnormal{#2}}
  }}}%
}

\begin{document}
\reportnum{-6}{DESY-23-087}
\title[Constraining the top-quark mass within the global MSHT PDF fit]{Constraining the top-quark mass within the global MSHT PDF fit}

\author[1]{Thomas Cridge}\email{thomas.cridge@desy.de}

\author[2]{Matthew A.~Lim}\email{m.a.lim@sussex.ac.uk}

\affil[1]{\orgdiv{Deutsches Elektronen-Synchrotron DESY}, \orgaddress{\street{Notkestr. 85}, \postcode{22607} \city{Hamburg}, \country{Germany}}}

\affil[2]{Department of Physics and Astronomy, University of Sussex, \orgaddress{\street{Sussex House}, \city{Brighton}, \postcode{BN1 9RH}, \country{UK}}}

\abstract{We examine the ability of experimental measurements of top-quark pair production to constrain both the top-quark mass and the strong coupling within the global MSHT parton distribution function (PDF) fit. Specifically, we consider ATLAS and CMS measurements of differential distributions taken at a centre-of-mass energy of 8 TeV, as well as $\unboldmath\( t\bar{t}\)$ total cross section data taken at a variety of experiments, and compare to theoretical predictions including next-to-next-to-leading order corrections. We find that supplementing the global fit with this additional information results in relatively strong constraints on the top-quark mass, and is also able to bound the strong coupling in a limited fashion. Our final result is $\unboldmath\( \mt=173.0\pm0.6~\gev \)$ and is compatible with the world average pole mass extracted  from cross section measurements of $\unboldmath\(172.5\pm0.7~\gev\)$ by the Particle Data Group. We also study the effect of different top-quark masses on the gluon parton distribution function, finding changes at high $x$ which nonetheless lie within the large PDF uncertainties in this region.  }

\maketitle
\clearpage

\tableofcontents

\clearpage

\section{Introduction}\label{intro}
The lack of any clear signals of New Physics from the experiments at the Large Hadron Collider (LHC) suggests the need to move towards precision measurements, with the aim to use these as a means to detect beyond the Standard Model effects indirectly. Of the Standard Model (SM) parameters, the top-quark mass is of significant importance, due to the strength of its coupling to the Higgs boson, the r\^ole it plays in governing the stability of the electroweak vacuum, and the fact that it is an important input to calculations of several backgrounds for important LHC processes.

Approaches to top-quark mass extractions typically follow one of two paths: either a direct reconstruction of the top-quark decay products is attempted, or the dependence of kinematic distributions and total cross sections on the mass is exploited. The former approach generally relies on modelling by Monte Carlo event generators, resulting in the determination of a so-called `Monte Carlo mass'. There remains some debate as to the interpretation of this quantity and whether it can be connected to a well-posed definition. The reliability of the latter method, meanwhile, is dependent upon the perturbative accuracy of the theoretical calculation which is used to extract the mass. It is also dependent, to a certain extent, on Monte Carlo modelling, which is used to extrapolate the measurement to the full phase space from the fiducial region. A comprehensive review of these issues can be found in ref.~\cite{Corcella:2019tgt}.

At the same time, the methodology of the fitting procedure must be carefully considered. It is important to bear in mind that, in addition to any dependence on SM parameters, any theoretical calculation at a hadron collider also implicitly relies on a number of parton distribution function (PDF) parameters which are normally extracted in separate fits. There may be significant correlations between the extracted SM parameters and the externally-fitted PDF parameters, which if not taken into account may bias the extracted result. This has been explicitly demonstrated in the case of extractions of the strong coupling, $\as$ \cite{Forte:2020pyp}, and consequently extractions of the strong coupling are typically performed within PDF collaborations \cite{Ball:2018iqk,Hou:2019efy,Cridge:2021qfd,H1:2021xxi,dEnterria:2022hzv}. The same considerations apply more generally \cite{Carrazza:2019sec, Greljo:2021kvv}, and hence also hold in the case of the top-quark mass.

Previous top-quark mass extractions have largely relied on measurements of the $t\bar{t}$ total cross section \cite{ATLAS:2014nxi,CMS:2016yys,CMS:2018fks}, utilising next-to-next-to-leading order (NNLO) theory predictions obtained via the program \texttt{top++}~\cite{Czakon:2011xx,Czakon:2013goa}. In addition, attempts to incorporate differential information for the $t\bar{t}$ process have been obtained using NLO theory predictions both alone~\cite{CDF:2014upy,ATLAS:2018fwq, CMS:2018tye,Gombas:2022zbh,ATLAS:2022jbw,ATLAS:2022jpn} and alongside a PDF fit~\cite{Guzzi:2014wia,CMS:2019esx}; indeed, the ability of the differential information to constrain $\mt$ in a global fit context was pointed out in ref.~\cite{Kadir:2020yml}, albeit in the context of error updating rather than full refitting. Extensions to include NNLO theory predictions have been performed in a joint fit with $\as$ (including the exact NNLO top-quark mass dependence)~\cite{Cooper-Sarkar:2020twv} or in a global fit (where the NNLO dependence for top-quark masses other than $\mt=173.3~\mathrm{GeV}$ is approximated through the use of NLO $K$-factors)~\cite{Gao:2022srd}. Finally, the single-top production process has been used to assess top-quark mass bounds~\cite{Alekhin:2016jjz,CMS:2017mpr}, in some cases using NNLO theory predictions~\cite{Gao:2020nhu}, although the process itself shows a reduced sensitivity. A summary of further top-quark mass measurements can be found in ref.~\cite{Schwienhorst:2022yqu}.

In this work, we combine calculations at the highest available order in the strong coupling for top-quark pair production (NNLO) with differential measurements from the ATLAS and CMS experiments taken at a centre-of-mass energy of 8 TeV~\cite{ATLAS:2015lsn,CMS:2015rld}. We utilise the MSHT global PDF fitting framework to simultaneously constrain the top-quark mass as well as the PDF parameters, thereby naturally incorporating any correlations between these quantities. 

The paper is organised as follows. In sec.~\ref{setup}, we detail the datasets which we use in this work as well as the theoretical inputs. We discuss the fitting procedure which we use to assess the top-quark mass sensitivity and to ultimately obtain best-fit values with associated uncertainties. In sec.~\ref{sensitivity} we present the resulting global fit in the $\as,\,\mt$ plane and comment on the individual distributions. Sec.~\ref{topmassMSHT} examines the constraints on the mass obtained using the default MSHT setup, while sec.~\ref{distributions} analyses the effects of different treatments of the available measured kinematic distributions. In sec.~\ref{alphas} we discuss the extent to which the top-quark datasets can contribute to constraints on the strong coupling, and in sec.~\ref{PDF} we illustrate the effect of the mass on the gluon PDF. We present a brief summary of our findings in sec.~\ref{summary} and finally conclude in sec.~\ref{conc}.

\section{Setup of the analysis}\label{setup}

\subsection{Experimental datasets} \label{data}
We consider the subset of top-quark measurements included in the MSHT20 PDF fit~\cite{Bailey:2020ooq} for which the NNLO QCD predictions at a variety of top-quark masses are currently available. We will therefore primarily focus on the single differential data in the lepton+jet channel at 8~TeV from ATLAS~\cite{ATLAS:2015lsn} and CMS~\cite{CMS:2015rld}. These data are each presented differentially in four distributions, namely the top-quark pair invariant mass, $\mtt$, the individual top-quark/antiquark transverse momentum, $\pT$, and the individual and pairwise rapidities, $\yt$ and $\ytt$. In MSHT, the absolute distributions are chosen in preference to the normalised, in order not to lose constraining information from the total cross-section integrated over bins.

For the case of the ATLAS data, the data is provided in both absolute and normalised form and so we choose the former. In addition, the full statistical correlations between distributions are provided~\cite{ATLAS:2015lsn,ATLAS:2018owm}, in principle allowing all four distributions to be fit simultaneously. In practice, this has been found to be very difficult by several groups~\cite{Czakon:2016olj,Bailey:2019yze,Thorne:2019mpt,Hou:2019efy,Amat:2019upj,Hou:2019gfw,Bailey:2020ooq,Kadir:2020yml,Amoroso:2020lgh,NNPDF:2021njg,Amoroso:2022eow}, and problems have been encountered not only in fitting the different distributions together but even in fitting the individual rapidity distributions. As a result, several different approaches have been taken to the inclusion of these data~\cite{Bailey:2019yze,Bailey:2020ooq,ATLAS:2018owm,Hou:2019efy,NNPDF:2021njg,Amoroso:2022eow}: here we will follow the MSHT20 approach and decorrelate the parton shower systematic between all four distributions and additionally into two components within each of the distributions. This is a conservative approach and we will analyse the effects of different choices of decorrelation and distribution in sec.~\ref{distributions}, where we will also elaborate further on our procedure. 

For the corresponding CMS data,  the statistical correlations between the distributions are not provided -- we therefore fit one distribution at a time, with our default choice being the top-antitop pair rapidity, as in MSHT20. We consider alternative choices of distribution in sec.~\ref{distributions}. All CMS measurements which we consider were originally presented as normalised distributions but have been converted to absolute distributions using the corresponding total cross section data~\cite{CMS:2016yys}. Further details on the inclusion of these datasets within MSHT20 are provided in refs.~\cite{Harland-Lang:2014zoa,Bailey:2019yze,Bailey:2020ooq}.

In this study we do not include data taken in the dilepton channel, e.g. the ATLAS single differential data at 7 and 8~TeV~\cite{ATLAS:2016pal} or the CMS double differential data~\cite{CMS:2017iqf}, even though they are included by default in MSHT20. This is because the NNLO theoretical predictions are only publicly available at a single top-quark mass. 

Whilst we focus our attention largely on the more constraining single differential measurements in the lepton+jet channel at 8~TeV, a number of measurements of the total $t\bar{t}$ cross section from ATLAS~\cite{ATLAS:2010zaw,ATLAS:2011whm,ATLAS:2012ptu,ATLAS:2012aa,ATLAS:2012xru,ATLAS:2012qtn,ATLAS:2015xtk} and CMS~\cite{CMS:2012ahl,CMS:2012exf,CMS:2012hcu,CMS:2013nie,CMS:2013yjt,CMS:2013hon,CMS:2014btv,CMS:2015auz} at 7~TeV and 8~TeV, and at the Tevatron \cite{CDF:2013hmv} were included in the MSHT20 PDF fit. We retain this (non-exhaustive) set of data in our analysis. We leave an analysis of the full set of top-quark total cross section data to a future study. 

\subsection{Theoretical predictions} \label{theory}
Throughout this work, we use theoretical predictions incorporating at least NNLO QCD corrections and work with the top-quark pole mass. For the total $t\bar{t}$ cross section we use the NLO QCD prediction from APPLGrid-MCFM~\cite{Carli:2010rw,Campbell:2012uf} and the NNLO prediction as implemented in \texttt{top++}, evaluated at the scale \mbox{$\mu_R=\mu_F=\mt$} ~\cite{Czakon:2011xx,Barnreuther:2012wtj,Czakon:2012zr,Czakon:2012pz,Czakon:2013goa}. Central predictions are made at a pole mass of $\mt = 172.5~\mathrm{GeV}$, and a variation of $1~\mathrm{GeV}$ in the mass is translated into a 3\% change in the cross section.\footnote{This dependence for the NLO cross section (also considering the effect of soft gluon resummation) was suggested in refs.~\cite{Catani:1996dj,Cacciari:2011hy}. We have verified it holds to within 0.1 percentage points for the NNLO cross section between $169-175~\gev$ using \texttt{top++}. Most of this dependence on $\mt$ is due to the overall $1/m_t^4$ dependence of the cross section, and therefore independent of the PDFs and $\as$.} The differential cross section for this process is known up to NNLO QCD~\cite{Catani:2019iny,Catani:2019hip} with NLO EW and resummation effects~\cite{Czakon:2017wor,Czakon:2019txp}. In this work we use the NNLO QCD predictions~\cite{Czakon:2015owf,Czakon:2016ckf,Czakon:2016dgf} computed using the \texttt{Stripper} framework~\cite{Czakon:2010td,Czakon:2014oma}. These are implemented via \texttt{fastNLO} tables~\cite{Kluge:2006xs,Britzger:2012bs,Czakon:2017dip}, which allow for rapid evaluations for a fixed set of observables and binnings. We use the same tables which were first made available in ref.~\cite{Cooper-Sarkar:2020twv}, with top-quark masses of \mbox{$\mt=\{169.0,171.0,172.5,173.3,175.0\}~\gev$} and which share the binnings of the experimental measurements. In addition, we supplement the NNLO QCD predictions with NLO electroweak (EW) $K$-factors~\cite{Czakon:2019txp}.

The renormalisation and factorisation scales chosen are those found to be optimal in ref.~\cite{Czakon:2016dgf}, namely,
\begin{align}
\mu_R&=\mu_F=H_T/4, \;\; \text{for} \;\; \mtt,\, \yt, \, \ytt \,, \\
\mu_R&=\mu_F=M_T/2, \;\; \text{for} \;\; \pT \,.
\end{align}

\subsection{The MSHT global fit} \label{methodology}
We use the MSHT global PDF fitting framework to combine the experimental measurements with the theoretical predictions. MSHT20~\cite{Bailey:2020ooq} is a global PDF fit utilising 4363 data points across 61 different datasets spanning older fixed target data, HERA deep inelastic scattering, neutrino dimuon, Tevatron and a wide range of recent LHC data to provide a state of the art determination of proton structure in terms of unpolarised proton PDFs. An extensive parameterisation of the PDFs at the input scale ($Q_0^2=1~\mathrm{GeV}^2$) incorporating 52 parton parameters is used to define the central values of the PDFs. The PDF uncertainties are defined via the Hessian method, with a 32-member subset of the parton parameters used to define an eigenvector basis. Given the global nature of the PDF dataset and the finite order at which the theory is implemented (NNLO QCD + NLO EW), rather than implementing a $\Delta\chi^2=1$ criterion to define the PDF uncertainties we use the so-called ``dynamic tolerance procedure''. This is more conservative and is motivated by a weaker hypothesis-testing criterion~\cite{Martin:2009iq} (see below in the context of the $\mt$ and $\as$ bounds in eq.~\ref{boundequation}); it ensures that the bounds on each eigenvector are such that each dataset sits within its 68\% confidence level limit.

We have implemented the NNLO QCD theoretical predictions (with NLO EW corrections where specified in sec.~\ref{theory}) within the MSHT20 global fit, and refit for each of the five available top-quark masses. In this way, we naturally account for correlations between the extracted PDFs, $\mt$ and $\as$. Performing this across a range of $\as$ and $\mt$ values we are then able to use the qualities of the different fits to analyse the sensitivity of the experimental data to the parameters of interest. After examining the two-dimensional dependence in the $\mt$-$\as$ plane, we will follow the procedure typically used in $\as$ extractions from PDFs (including in the most recent MSHT20 determination of the strong coupling~\cite{Cridge:2021qfd}) to consider the extent to which bounds can be placed.  We outline this procedure here. 

We consider the variation in the $\chi^2_n$ for the $n$-th dataset with $N$ degrees of freedom (data points) as we scan along a particular parameter or eigenvector direction, and assume that it follows a $\chi^2$ distribution, i.e. 

\begin{equation}
    P_N(\chi^2) = \frac{(\chi^2)^{N/2-1}\exp{(-\chi^2/2)}}{2^{N/2}\Gamma(N/2)}\,.
\end{equation}

We can then obtain the $m$-th percentile, $\xi_m$ by solving:

\begin{equation}
    \int_0^{\xi_m} \mathrm{d}\chi^2 P_N (\chi^2) = m/100
\end{equation}

Then $\xi_{50} \approx N$ is the most probable value, and $\xi_{68}$  will be used in the definition of the 68\% confidence level uncertainties. The ratio $\xi_{68}/\xi_{50}$ will then reduce with the number of data points $N$.

In order to define the 68\% confidence limit on a quantity ($m_t$ or $\as$), or indeed on a PDF eigenvector, from this particular dataset about the global minimum $\chi^2_{n,0}$, we then choose to set the bounds by the condition:

\begin{equation} \label{boundequation}
    \chi_n^2 \lt \frac{\chi^2_{n,0}}{\xi_{50}}\xi_{68},
\end{equation}

where $\chi^2_{n,0}$ is the $\chi^2$ of the dataset at the global minimum.

This effectively rescales the $\chi^2_n$ for this dataset up by a factor of $\xi_{68}/\xi_{50}$ and accounts for the fact that the $\chi^2$ of the dataset at the global minimum is likely to be different from that of the particular dataset. Alternatively, it can be equivalently understood as a rescaling of the 68\textsuperscript{th} percentile $\xi_{68}$ by $\chi^2_{n,0}/\xi_{50}$. As the ratio $\xi_{68}/\xi_{50}$ reduces with the number of data points $N$, this also reflects the intuition that datasets with more data points $N$ are likely to sit closer to the global minimum and so require a smaller rescaling of the uncertainty condition. Once this condition is exceeded in the scan along the parameter value or eigenvector direction, we interpret this as a bound. We interpolate between this and the penultimate point to obtain the precise value of the bound.

This procedure is repeated for all of the datasets in the fit, and in order to ensure that each dataset lies within its 68\% confidence level (as defined by eq.~\ref{boundequation}), the most stringent of the bounds is then taken. The dataset corresponding to said bound is then interpreted as constraining the quantity of interest.

This is the methodology used within MSHT20 to define the uncertainties on each PDF eigenvector, and as a consequence on the PDFs themselves. It can be extended to consider any further parameter by fitting it together with the PDFs (and so account for correlations). This has been used on several occasions to provide bounds on $\as$ \cite{Harland-Lang:2015nxa,Cridge:2021qfd}, and we extend this here to consider $\mt$.

\section{Sensitivity of $t\bar{t}$ distributions to $\mt$ and $\as$}\label{sensitivity}
\begin{figure}[ht!]
\centering
\includegraphics[width=0.75\textwidth]
{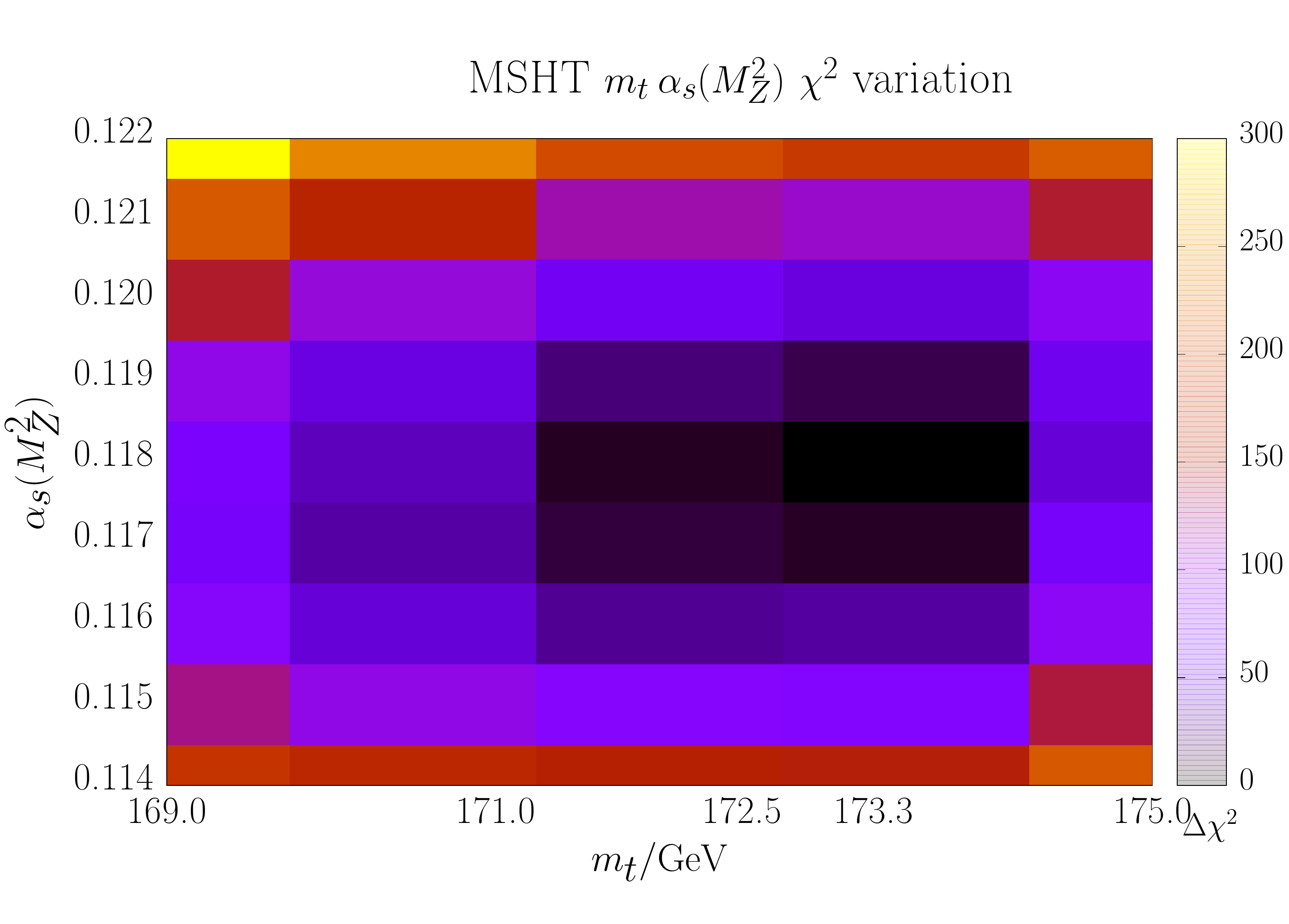}
\caption{Heat map showing the minimum $\chi^2$ value obtained from fits with varying $\mt$ and $\as$. All datasets are included, as described in sec~\ref{data}.}
\label{heatmapalldata}
\end{figure}

\begin{figure}[ht!]
\centering
\includegraphics[width=0.49\textwidth]
{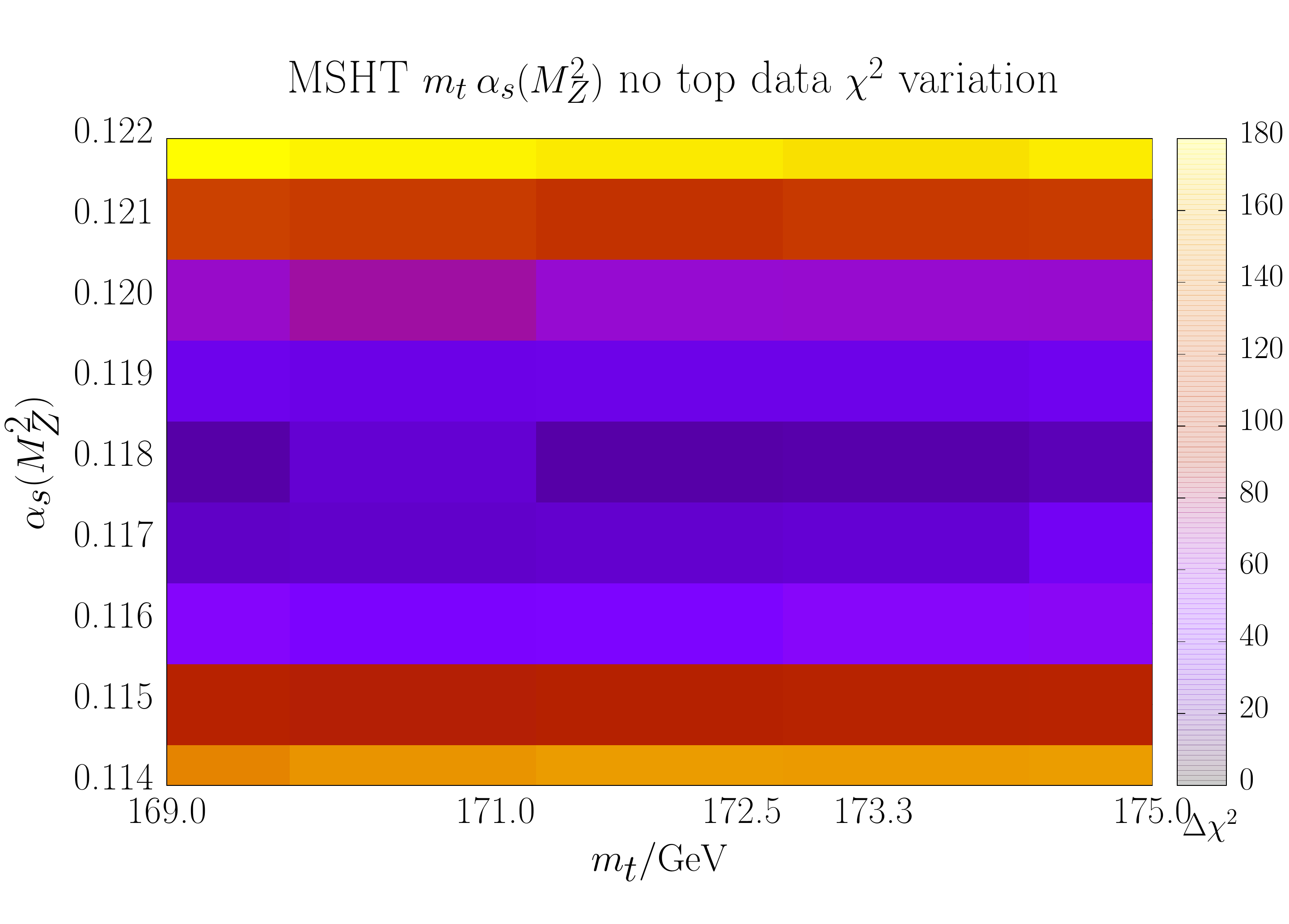}
\includegraphics[width=0.49\textwidth]{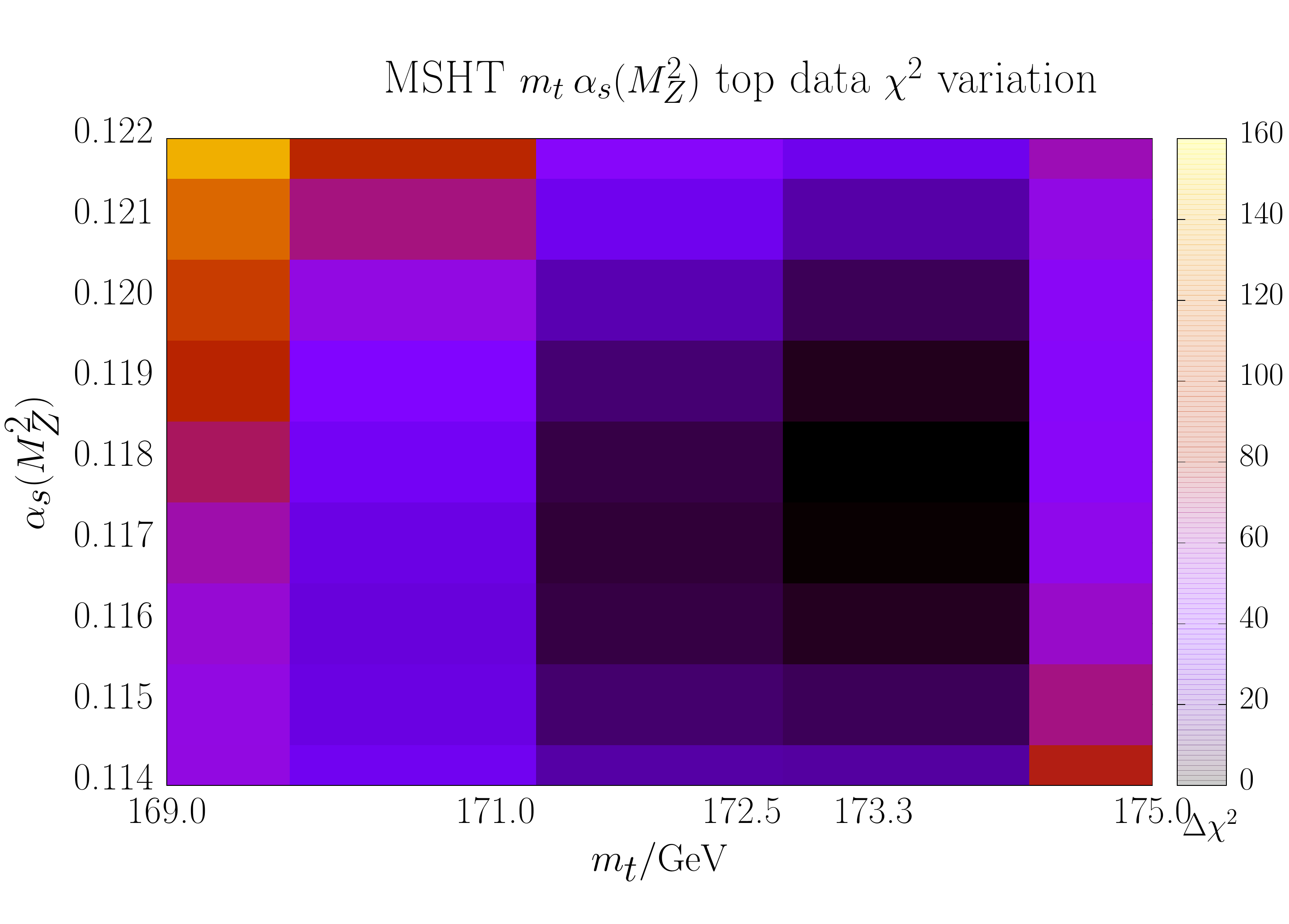}
\caption{Heat map showing the minimum $\chi^2$ value obtained from fits with varying $\mt$ and $\as$. Left: top-quark datasets removed. Right: top-quark data only. Note that these are the same fits as those shown in fig.~\ref{heatmapalldata}.}
\label{heatmapnotopalltop}
\end{figure}

Before performing any parameter extraction, it is instructive to examine the two-dimensional dependence of the global fit quality on $\mt$ and $\as$. To that end, we perform fits 
for the five values of the top-quark mass available to us and with 9 equally-spaced $\as$ values from 0.114 to 0.122. We present the results of the global fits in fig.~\ref{heatmapalldata}. We observe that we are able to constrain both parameters simultaneously, with the fit finding a global minimum for $\mt\sim 173.3~\gev,\,\as\sim 0.118$. We do not observe any clear signs of degeneracy, which would indicate significant correlation between the parameters. In fig.~\ref{heatmapnotopalltop} we examine the impact of  the subset of the global fit dataset corresponding to top-quark measurements. We show the heatmap from the same fits as in fig~\ref{heatmapalldata}, but with top-quark data removed, as well as the corresponding plot including only the top-quark data. We observe that the former plot is unable to constrain the top-quark mass, while the latter shows a similar pattern in $m_t$ to that observed in fig~\ref{heatmapalldata}, but displays a weaker dependence on $\as$. This indicates the importance of the top-quark datasets in constraining $\mt$, while the remainder of the global dataset offers stronger constraints on $\as$.

\begin{figure}[ht!]
\centering
\includegraphics[width=0.49\textwidth]{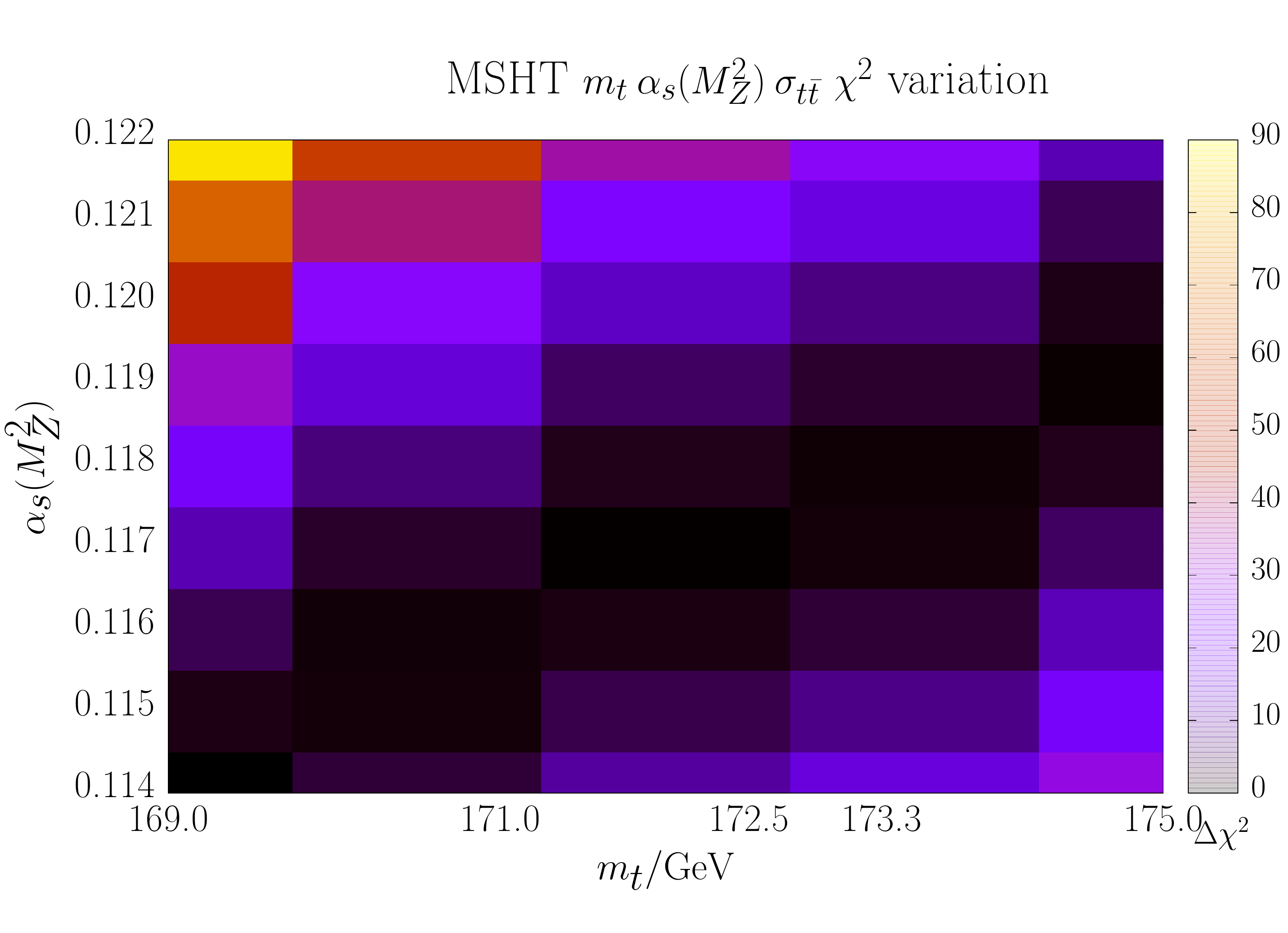}\\
\includegraphics[width=0.49\textwidth]{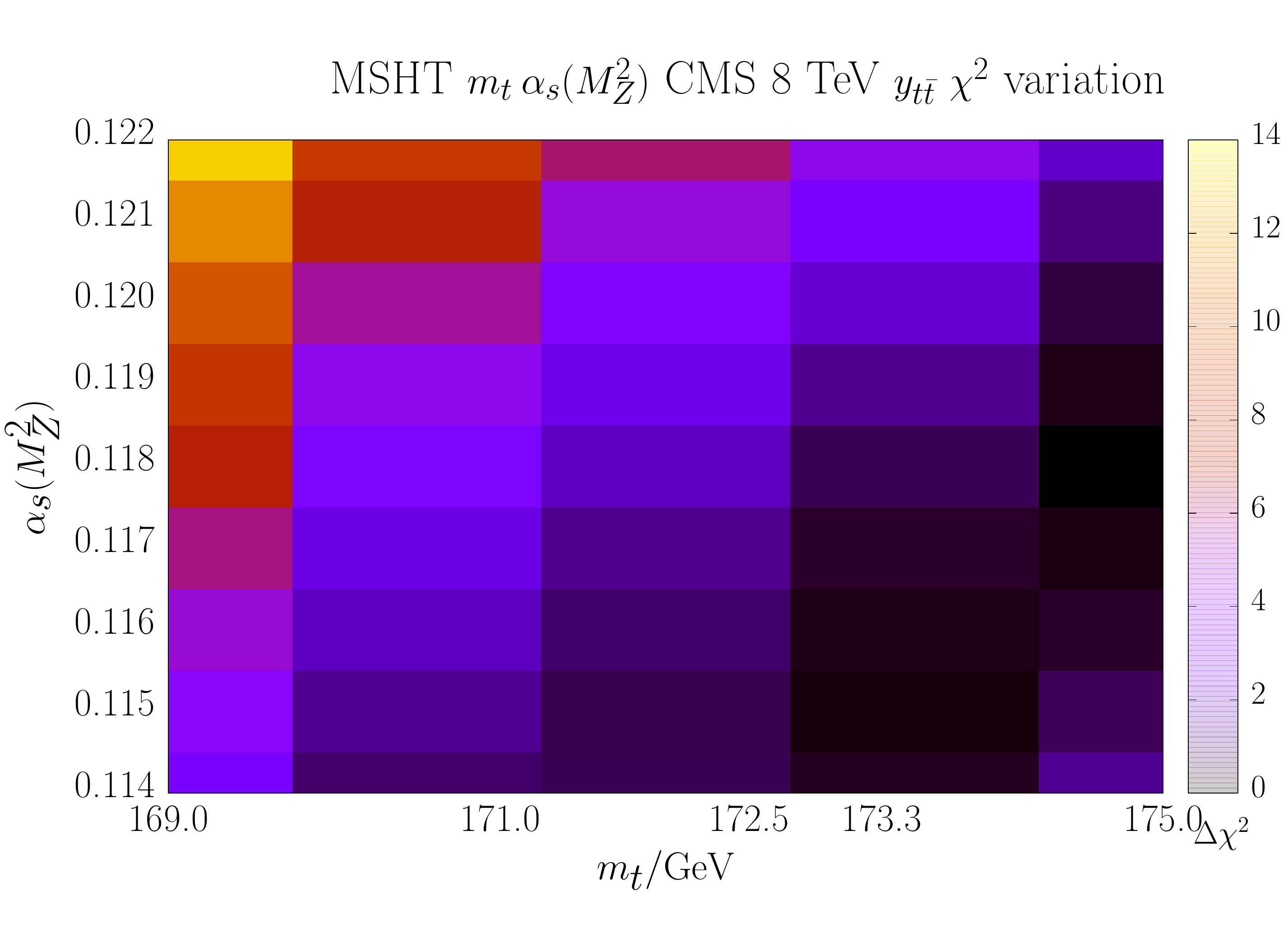}
\includegraphics[width=0.49\textwidth]{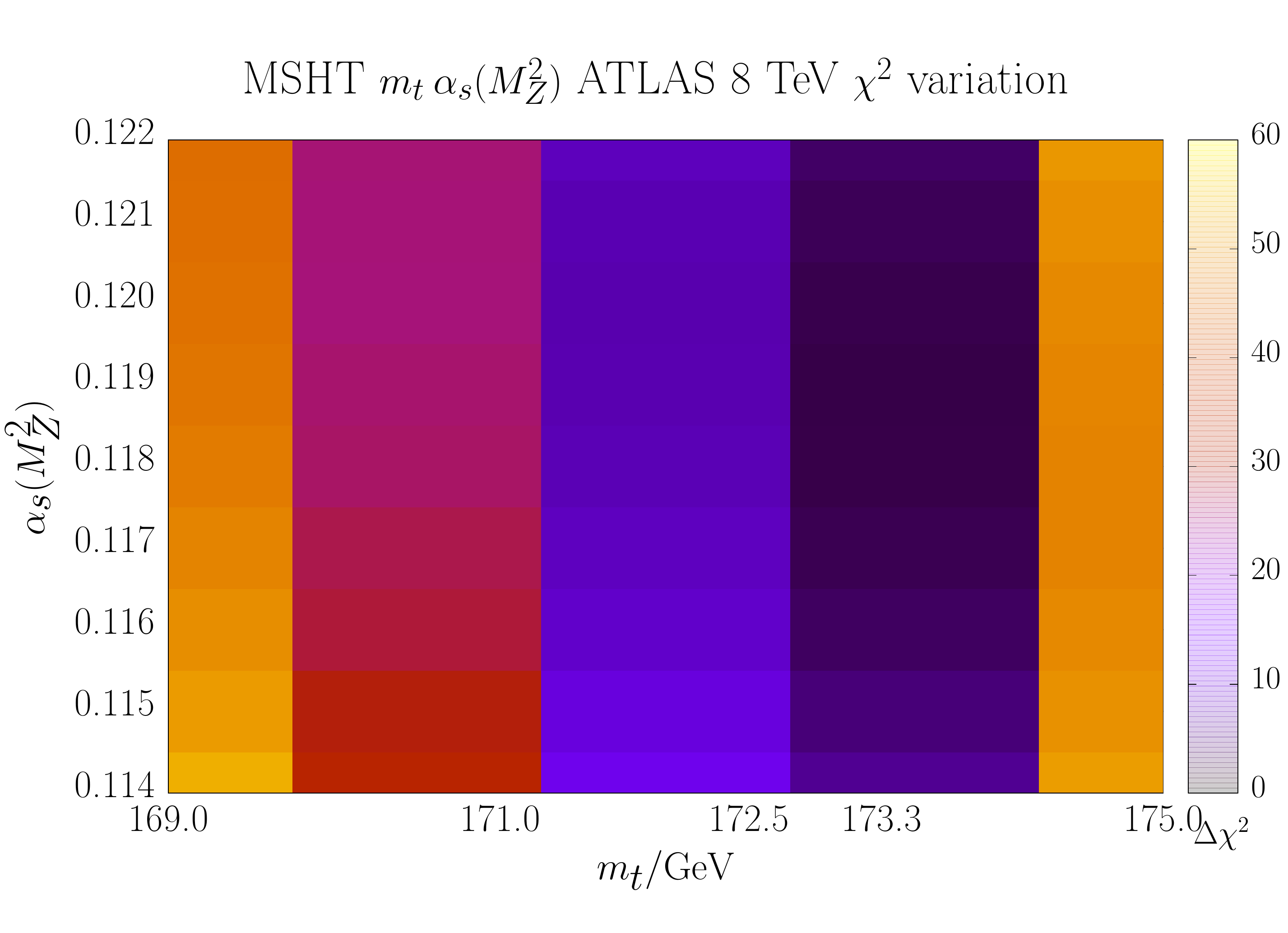}
\caption{Heat map showing the minimum $\chi^2$ value obtained from fits with varying $\mt$ and $\as$. Top: $\ttbar$ total cross section only. Lower left: CMS $\ytt$ only. Lower right: ATLAS $\pT,\,\mtt,\,\yt,\,\ytt$. Note that these fits are identical to those shown in fig.~\ref{heatmapalldata}.}
\label{heatmapdatasets}
\end{figure}

We turn to an examination of the individual top-quark datasets. In fig.~\ref{heatmapdatasets} we show again the same fits as in figs.~\ref{heatmapalldata} and \ref{heatmapnotopalltop}, but this time separating out the contributions from the total $\ttbar$ cross section data, the ATLAS multi-differential data and the single-differential CMS data in $\ytt$ (the MSHT20 default), the latter two both measured in the lepton+jets channel at 8~TeV. We notice a clear degeneracy in the total cross section data, indicative of the well-known fact that this alone is unable to constrain both $\mt$ and $\as$ due to the compensatory nature of the joint dependence. The CMS data seem to offer slightly improved constraining power, although signs of degeneracy are still present. The ATLAS data, in contrast, provide very strong constraints on $\mt$ while being much more weakly constraining in the $\as$ direction. This is likely to be due to the fact that while the ATLAS data contain measurements of all four kinematic distributions, the included CMS data are taken from a rapidity distribution, which one expects on theoretical grounds to be naturally less strongly dependent on the top-quark mass.

\begin{table}[ht!]
\begin{center}
\renewcommand\arraystretch{1.25}
\begin{tabular}{|l|l|l|l|}
\hline
   $\mt$ (GeV)       &  $\chi^2_{\rm global}$  &   $\chi^2_{\rm top}$ &  
$\as$        \\
          &  4363 pts &  51 pts &  \\
\hline
169.0   &  5152.7  & 116.5   & 0.1170     \\
171.0   &  5117.4  & 84.3   & 0.1173     \\
172.5   &  5091.5 & 59.4   & 0.1175     \\
173.3  & 5084.9  & 52.5  & 0.1175   \\
175.0  & 5129.8  &  96.0  &  0.1178  \\
\hline
    \end{tabular}
\end{center}

\caption{The quality of the fit as a function of the top-quark mass $\mt$ with $\as$
left free.}
\label{mtasfree}   
\end{table}

Finally, in tab.~\ref{mtasfree} we examine the best-fit $\as$ value obtained in the fit for the five different $\mt$ values, as well as the global and top-quark data total $\chi^2$ values. One can again see the preference for $\mt \sim 173.3~\gev$, with a corresponding best fit $\as=0.1175$. Moreover, it can be observed that the best fit $\as$ does not vary significantly with $\mt$ -- indeed the whole range shown here is within the uncertainties of the result quoted in ref.~\cite{Cridge:2021qfd}, whilst the best fit is very close to the best fit of 0.1174 obtained there. Whilst $\as$ does increase slightly with $\mt$, likely to counter-balance the effect of reducing the cross-section with increasing $\mt$, the change is relatively small.

The results in this section therefore strongly suggest that the correlation between $\mt$ and $\as$, in the context of the global fit which we perform, is limited,  at least in the vicinity of the best fit minimum. Indeed, both fig.~\ref{heatmapalldata} and tab.~\ref{mtasfree} imply that the $\as$-$\mt$ dependence is quasi-one-dimensional in the fit as a whole. We will exploit this property in order to constrain individually the two parameters, following the method described in sec.~\ref{methodology}. 

\section{Constraining the top-quark mass within the MSHT default setup}\label{topmassMSHT}
Given the findings of sec.~\ref{sensitivity}, in this section we proceed with a one-dimensional extraction of $\mt$ at a fixed value of $\as=0.118$. Following the methodology described in sec.~\ref{methodology}, we interpolate the $\chi^2$ dependence of the fit as a function of $\mt$ and assume that it follows a cubic dependence about its minimum. We have found a cubic function to be necessary to describe in particular the ATLAS dataset, i.e. we include the next term in the Taylor expansion of the $\chi^2$ function about its minimum. This allows us to include all data points in $\mt$, even when some of these are relatively far from the minimum for this strongly constraining dataset. As discussed previously, we adopt a more conservative tolerance-based definition of the $\Delta\chi^2$ in order to set limits on the parameters, rather than using a simple $\Delta\chi^2=1$ criterion.

\begin{figure}[ht!]
\centering
\includegraphics[width=0.75\textwidth]{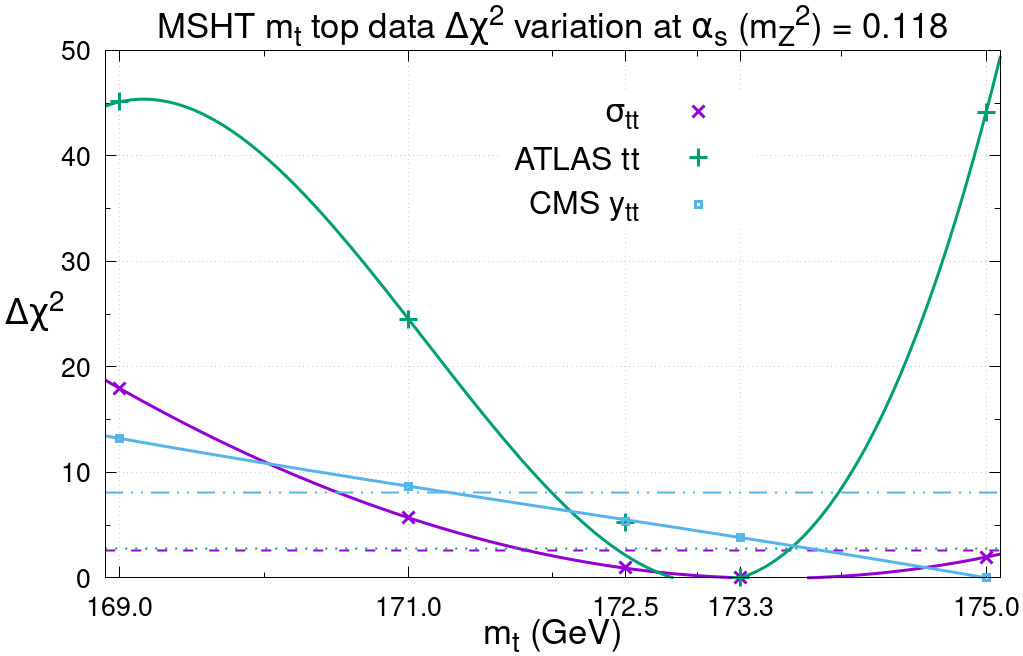}
\caption{$\Delta\chi^2$ of the included top-quark datasets in the MSHT default setup (see sec.~\ref{data}) as a function of $\mt$ and with fixed $\as=0.118$.}
\label{mtscandefault}
\end{figure}

In fig.~\ref{mtscandefault} we show the dependence of the $\chi^2$ values for various top-quark datasets as a function of $\mt$ at fixed $\as=0.118$. In particular, we compare the constraining power of the total cross section data, the ATLAS multi-differential data and the CMS pair-rapidity distribution $\ytt$ in the same fit. We remind the reader that this combination of distributions defines the MSHT default setup, albeit omitting the ATLAS and CMS dilepton datasets. The horizontal lines represent the bounds on $\Delta\chi^2_n$ of the three top-quark datasets from the global minimum (c.f. eq.~\ref{boundequation}). 
	
We observe that the CMS $\ytt$ distribution provides only a one-sided bound on $\mt$ (favouring high values) over the region of $\mt$ sampled\footnote{We remind the reader that a one-sided bound implies that the dataset sits away from its minimum in the global fit, i.e. it is in some tension with other datasets.}, reflecting the limited sensitivity of this distribution. This is in accordance with fig.~\ref{heatmapdatasets} and with observations in ref.~\cite{Cooper-Sarkar:2020twv}. We find a lower bound of $\mt\sim 171.3~\mathrm{GeV}$. Next we consider the total cross section $\sigma_{\ttbar}$. We note that this dataset is approximately locally quadratic about the global minimum, and hence is able to provide a two-sided constraint on the mass. Whilst we find a relatively weak upper bound ($\sim 175.2~\gev$), the lower ($\sim 171.8~\gev$) is slightly stronger than the CMS $\ytt$ distribution. Finally, the ATLAS multi-differential data show a greater sensitivity to  $\mt$, again providing a two-sided bound $172.4~\gev\lt\mt\lt173.6~\gev$. Taking the tightest bounds from all datasets considered, we find that the ATLAS dataset provides both the upper and lower values.

\begin{figure}[ht!]
\centering
\includegraphics[width=0.49\textwidth]{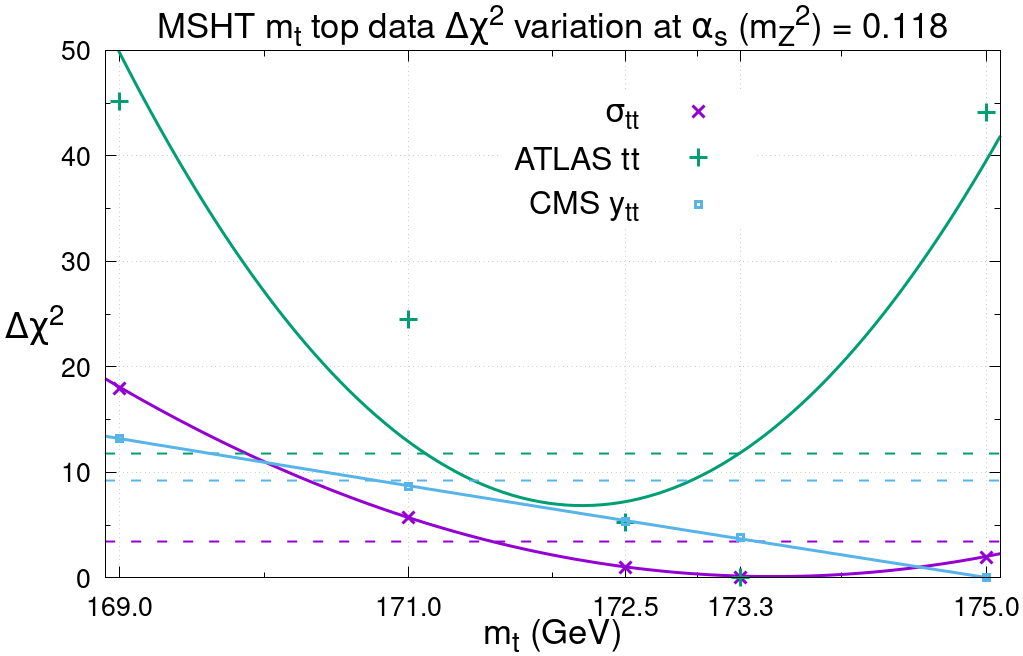}
\includegraphics[width=0.49\textwidth]{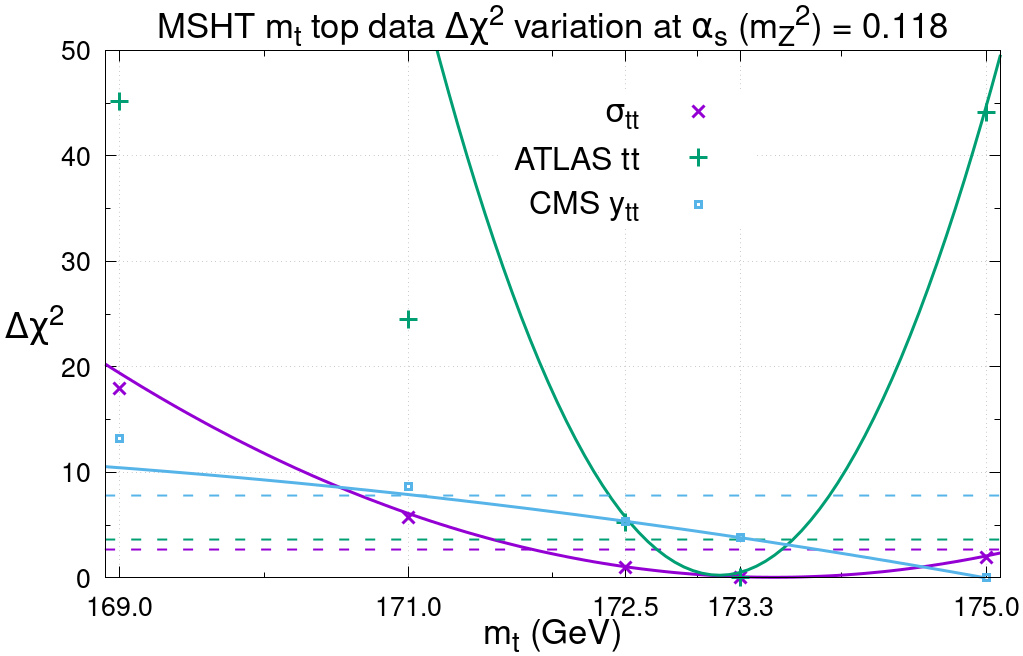}
\includegraphics[width=0.49\textwidth]{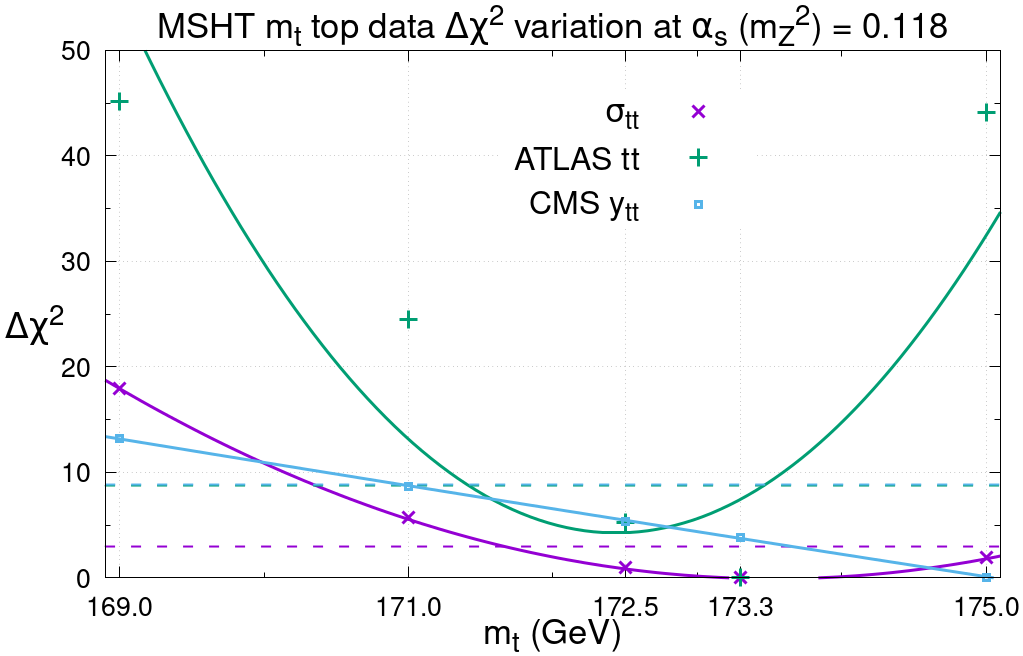}
\caption{$\Delta\chi^2$ of our default included top-quark datasets (see sec.~\ref{data}), but with alternative interpolation procedures as a function of $\mt$ and with fixed $\as=0.118$. Upper left: quadratic fit, all points equally weighted. Upper right: quadratic fit, neglecting first two $\mt$ points. Lower: quadratic fit, up-weighting points closest to the minimum.}
\label{mtscanaltinterpolation}
\end{figure}

Before moving on, we assess how our choice of interpolation affects the bounds on $\mt$ which we are able to set. In fig.~\ref{mtscanaltinterpolation}, we consider three alternative options which are all based on a quadratic $\mt$ dependence: first, assigning all five points an equal weight; second, discarding the first two points (which are furthest from the global minimum); third, weighting the points to favour those closer to the global minimum. In the last case, we have considered several different possibilities for the weights -- we show in the figure a representative case where the points have been assigned relative weights of $\{1,2,3,5,1\}$.

We begin with the case of providing an equal weight to all points with a quadratic polynomial in fig~\ref{mtscanaltinterpolation}~(upper left). This is observed to work well for the total cross section and CMS datasets, causing the net lower and upper bounds across both datasets to change by $\sim 0.2-0.3~\gev$ and $\sim 0.5~\gev$ respectively. Nonetheless, as these data provide neither our most stringent upper nor lower bounds on $\mt$ this has no global effect. The ATLAS data provided our most constraining limits and thus the effects on this dataset are more important for the extraction of the $\mt$ bounds. However, assigning all points an equal weight in a quadratic fit clearly provides a very poor description of the ATLAS data -- sufficiently poor, indeed, that the bounds would be meaningless.

On the other hand, considering the second case of dropping the first two points in $\mt$ in fig~\ref{mtscanaltinterpolation}~(upper right), the ATLAS data is much better locally described by a quadratic function. Indeed, the bounds obtained would then be substantially stronger due to the tighter nature of the profile around the minimum, with the lower bound increasing by $\sim 0.2~\gev$. Nevertheless, following this procedure causes difficulty in extracting a bound for the CMS dataset, given its limited sensitivity. In this case, the total cross section data bounds are unaffected relative to our default cubic option. 

Finally, the intermediate option of up-weighting points closest to the minimum shown in  fig~\ref{mtscanaltinterpolation}~(lower) improves the situation relative to the equally-weighted case, whilst also removing the need to drop information from the first two $\mt$ values. In this case the bounds from the 
total cross section data shift by only $\sim 0.1~\gev$  relative to the better cubic fit. Once more, however, issues with the more constraining ATLAS dataset in particular remain. 

Overall, whilst we see slight changes in the bounds for the total cross section and CMS $\ytt$ data depending on the interpolation chosen, for the ATLAS dataset (which ultimately provides our most stringent bounds overall) only two forms produce reasonable fits in the vicinity of the $\mt$ minimum - the default cubic and the quadratic using only the last three $\mt$ values. The former has the advantage of using all the $\mt$ information available and additionally provides the more conservative bound. Therefore we note that whilst there is some uncertainty due to the exact interpolation performed, we justify our decision to utilise the cubic fit as our default on the basis that it provides more conservative bounds. In fact, the tighter bound which could be obtained from the ATLAS data is encompassed within the looser default range from the cubic fit: our default bounds therefore also include to some extent the effects of changing the interpolation. The difficulties we encounter in using a quadratic form for the $\mt$ dependence further motivate our cubic fit -- since the ATLAS dataset appears so strongly constraining, the extreme value $\mt=169.0~\gev$ therefore lies some distance from the minimum and it is not surprising that higher terms in the Taylor expansion of the $\chi^2$ function are needed to properly describe this region. 

\begin{figure}[ht!]
\centering
\includegraphics[width=0.9\textwidth]{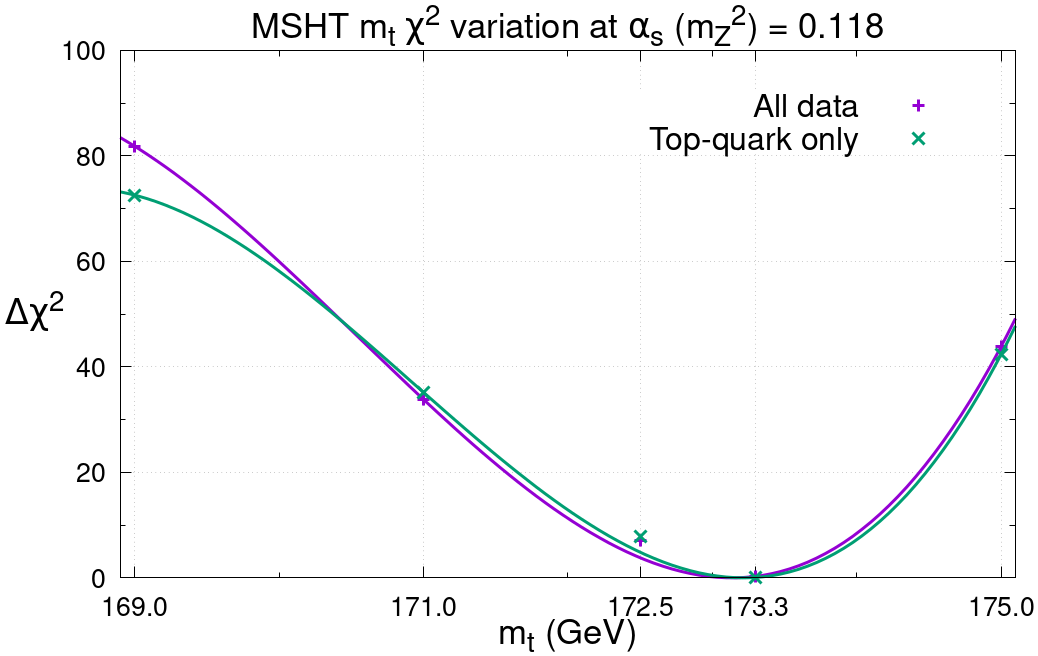}
\caption{$\Delta\chi^2$ profiles for the baseline global fit as $\mt$ is changed. $\Delta \chi^2$ for the global dataset and the top-quark data only are shown.}
\label{mtscanprofiles}
\end{figure}

Finally, in fig.~\ref{mtscanprofiles} we plot the $\Delta \chi^2$ as the top-quark mass is changed across the five different fixed values, again using a cubic interpolation. Both the change in $\chi^2$ for the total global fit and for the top-quark data are shown on the same scale, again demonstrating that the top-quark data contributes the overwhelming majority of the $\mt$ dependence to the PDF fit. We can also use these profiles to determine the $\Delta \chi^2$ of the global fit corresponding to our $\mt$ bounds: we find these correspond to $\Delta \chi^2 =\{3.2,\, 4.1\}$ for the lower and upper bounds respectively\footnote{This is smaller than the average of $\Delta\chi^2 \approx 10$ found for the PDF eigenvectors in the MSHT20 NNLO PDF fit, but well within the range of observed values.}. On the other hand, if we had taken the less conservative approach of using the $\Delta \chi^2=1$ criterion rather than the default MSHT dynamic tolerance then more stringent bounds would be obtained, viz. $172.7~\gev < \mt < 173.3~\gev$. However the usual issues of dataset tensions, methodological limitations and the finite-order nature of the theoretical predictions amongst other effects mean that the textbook scenario does not apply. Instead, in a global PDF fit the tolerance is used to account for these considerations, thus enlarging the uncertainties and providing our bounds.

\section{Assessing alternative treatments of the CMS and ATLAS data}\label{distributions}
In this section, we consider alternative possibilities for the inclusion of the differential CMS and ATLAS top-quark data in the lepton+jets channel, which differ from the default MSHT20 treatment~\cite{Bailey:2019yze}. Specifically, in the case of CMS we assess the options for the included distribution, while in the ATLAS case this picture is somewhat complicated by the statistical correlations between distributions. We therefore investigate the effect of different decorrelation treatments, which are themselves tied to the choice of distributions. 

\subsection{Choices of CMS distribution}\label{CMSdistribution}
Our comparison of the different datasets in sec.~\ref{topmassMSHT} revealed that the ATLAS multi-differential data were able to provide a significantly stronger constraint on $\mt$ than the CMS $\ytt$ distribution or the total cross section data alone. This is to some extent expected, since the availability of the statistical correlations between kinematic distributions enables the simultaneous inclusion of the $\pT,\,\mtt,\,\yt$ and $\ytt$ data in the ATLAS case. In contrast, this information is not publicly available in the CMS case and we are therefore only able to fit a single distribution at a time. In this section, we investigate different choices for this distribution compared to our default choice of $\ytt$. In all cases complete refits are done with the alternative CMS distribution.

\begin{figure}[ht!]
\centering
\includegraphics[width=0.49\textwidth]{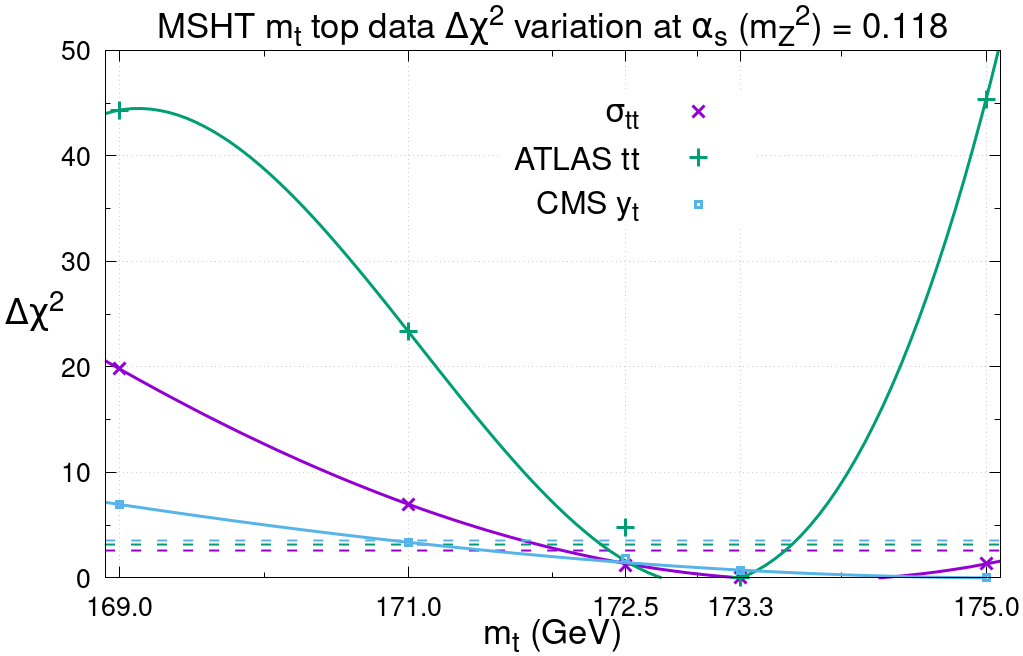}
\includegraphics[width=0.49\textwidth]{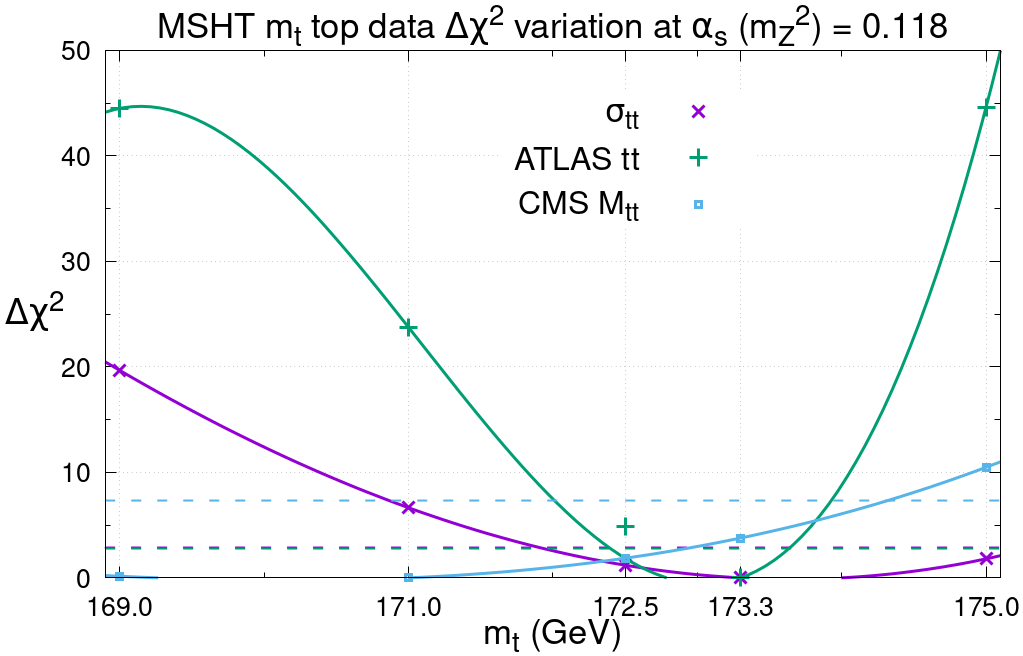}
\includegraphics[width=0.49\textwidth]{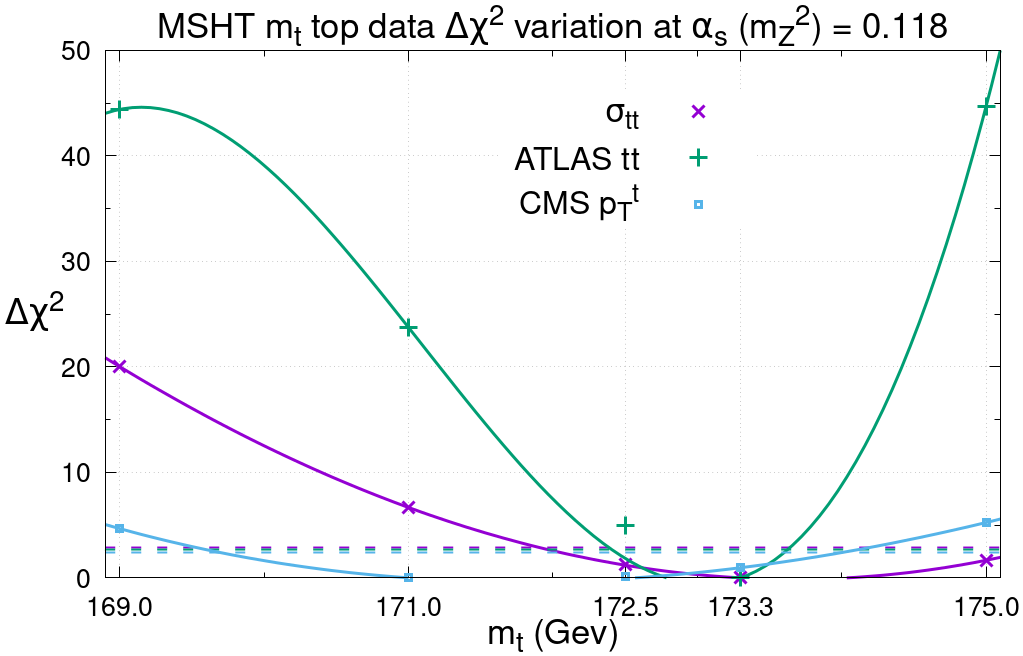}
\caption{$\Delta\chi^2$ of the default ATLAS and total cross section datasets, but with alternative choices of CMS distribution relative to our default. Plots are again as a function of $\mt$ and with fixed $\as=0.118$. Upper left: CMS single rapidity $\yt$. Upper right: CMS top-quark pair invariant mass $\mtt$. Lower: CMS top-quark transverse momentum $\pT$.}
\label{CMSchoices}
\end{figure}

Turning first to the case of the CMS $\yt$ distribution (fig.~\ref{CMSchoices}, upper left), we see that we can still only obtain a one-sided bound over the region of $\mt$ sampled. In general, the behaviour is very similar to the $\ytt$ case, albeit showing a slightly more quadratic dependence. The $\mtt$ distribution (fig.~\ref{CMSchoices}, upper right) instead shows a distinctly different behaviour, again providing only a reasonable one-sided bound but in this case at high, rather than low, $\mt$. Finally, in contrast the $\pT$ case (fig.~\ref{CMSchoices}, lower panel) shows a quadratic behaviour in the vicinity of the global minimum and sets limits $169.7~\gev\lt\mt\lt174.0~\gev$. Of the four cases, this choice provides the greatest sensitivity but remains notably worse than the ATLAS data, as observed in other studies -- for comparison, the corresponding ATLAS limits are $172.4~\gev<\mt<173.6~\gev$ (which remain the same regardless of the CMS distribution included in the fit), while the total cross section provides $171.9~\gev<\mt<175.4~\gev$ (which varies by at most $0.2~\gev$ on the upper and lower bounds as the CMS distribution is altered). The fact that the CMS $\mtt$ and $\pT$ distributions are generally better able to constrain $\mt$ is expected based on theoretical grounds~\cite{Czakon:2016vfr,Czakon:2016olj,Ju:2019mqc} and has also been demonstrated in e.g. ref.~\cite{Cooper-Sarkar:2020twv}\footnote{In fact, the choice of the $\ytt$ distribution in cases where correlations are unavailable is often motivated by the reduced sensitivity to the top-quark mass (which was beneficial before different $\mt$ theoretical predictions were available), as well as the reduced sensitivity to BSM and higher order effects \cite{NNPDF:2017mvq,NNPDF:2021njg,Kadir:2020yml}.}. Nonetheless, in all cases the CMS bounds are weaker than the corresponding ATLAS bounds.

\subsection{Choices of ATLAS kinematic distributions and decorrelation models}
In order to fit multiple ATLAS distributions simultaneously, it is necessary to include information about both the statistical and systematic correlations within and between the different kinematic variables. As remarked in sec.~\ref{data}, using the correlation matrices as provided by the experimental collaboration leads to very poor fit qualities (large $\chi^2/N$). The default MSHT20 procedure for dealing with this issue was detailed in refs.~\cite{Bailey:2019yze,Bailey:2020ooq} and entails decorrelating the two-point parton shower systematic across the four distributions. In addition, we make the conservative choice of further separating this source of uncertainty within the individual distributions into two pieces according to a trigonometric decomposition. The evaluation of this two-point systematic involves using the difference between two Monte Carlos to define a systematic uncertainty, which is taken as fully correlated across all bins. In reality, the correlations on these systematic uncertainties are not well known and therefore different levels of correlations can in principle be used. This was first investigated in the context of inclusive jet data from ATLAS~\cite{ATLAS:2017kux}. The assumption of full correlation across all bins is a strong one and in fact several studies have shown that by applying a small degree of decorrelation across the bins the fit quality can be improved significantly, see e.g. refs~\cite{Bailey:2019yze,Bailey:2020ooq}.

It has been observed~\cite{Bailey:2019yze,Hou:2019gfw,Amat:2019upj,Kadir:2020yml,Amoroso:2022eow} that, while it is possible to fit the $\pT$ and $\mtt$ distributions simultaneously by following the first part of the above prescription alone, as soon as rapidities are included the second part also becomes necessary. We therefore begin by simply fitting the $\pT$ and $\mtt$ distributions, decorrelating only between the distributions and not within. Given the findings of sec.~\ref{CMSdistribution}, we do this using the CMS $\pT$ distribution, which was found to be the most constraining of the four options, rather than using the MSHT20 default of the CMS $\ytt$. We remind the reader, however, that the precise choice of CMS distribution was found to have a negligible impact on the ATLAS bounds.

\begin{figure}[ht!]
\centering
\includegraphics[width=0.49\textwidth]{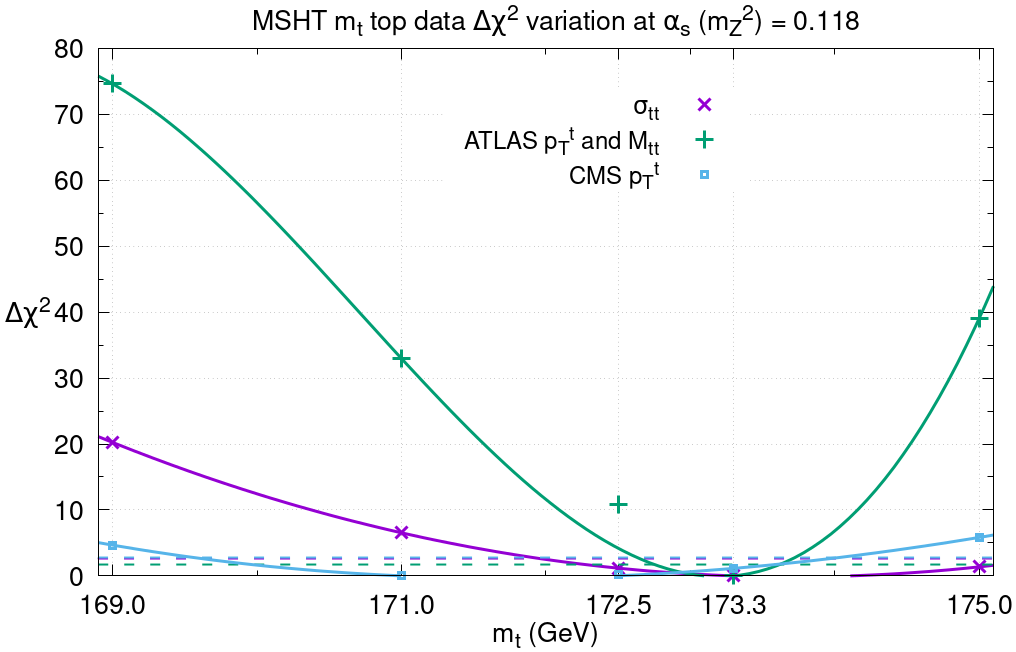}
\includegraphics[width=0.49\textwidth]{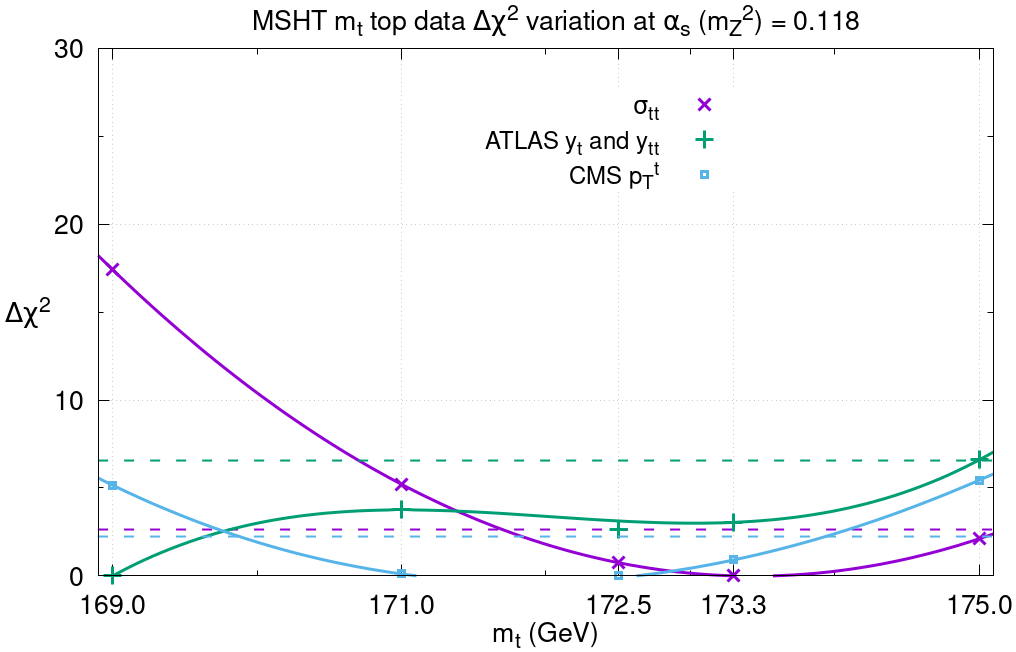}
\caption{$\Delta\chi^2$ of the ATLAS datasets (left) $\pT$ and $\mtt$ and (right) $\yt$ and $\ytt$, along with those of the CMS $\pT$ data and the total cross section data as a function of $\mt$ and with fixed $\as=0.118$.}
\label{ATLASptmttorytytt}
\end{figure}

We show the results in fig.~\ref{ATLASptmttorytytt}~(left) where again complete refits are performed with the new ATLAS distributions -- we note that the total cross section and CMS curves are largely unaffected by this change, while the ATLAS data become slightly less quadratic in the region near the minimum. With respect to the lower panel of fig.~\ref{CMSchoices}, the upper bound is unchanged while the lower bound becomes more stringent. This further demonstrates the conservative nature of our final uncertainty estimate on $\mt$ and the robustness of our procedure. We remark that this choice of treatment of the ATLAS data (i.e. including just $\pT$ and $\mtt$ and only decorrelating between distributions) is similar to that followed by the CT18 global PDF fit~\cite{Hou:2019efy}. We could instead take a choice similar to that made by the NNPDF4.0 global PDF fit~\cite{NNPDF:2021njg} and include both the single and pair rapidity distributions $\yt$ and $\ytt$, with the parton shower systematic then decorrelated between the two distributions but not within. Doing so, we observe poor fit qualities as expected ($\sim 2-3$ per point, similar to but actually lower than in the NNPDF4.0 study), but for the sake of completeness we still examine the extent to which the top-quark mass can be constrained in this case. Fig.~\ref{ATLASptmttorytytt}~(right) shows the $\chi^2$ profiles observed for the top-quark data. The reduced sensitivity of the ATLAS data to the top-quark mass is immediately clear, with the shape now not quadratic and rather flat in the vicinity of the global minimum $\mt$. This further attests to the fact that the constraints on $\mt$ in the ATLAS data arise predominantly from the $\pT$ and $\mtt$ distributions.

\begin{figure}[ht!]
\centering
\includegraphics[width=0.49\textwidth]{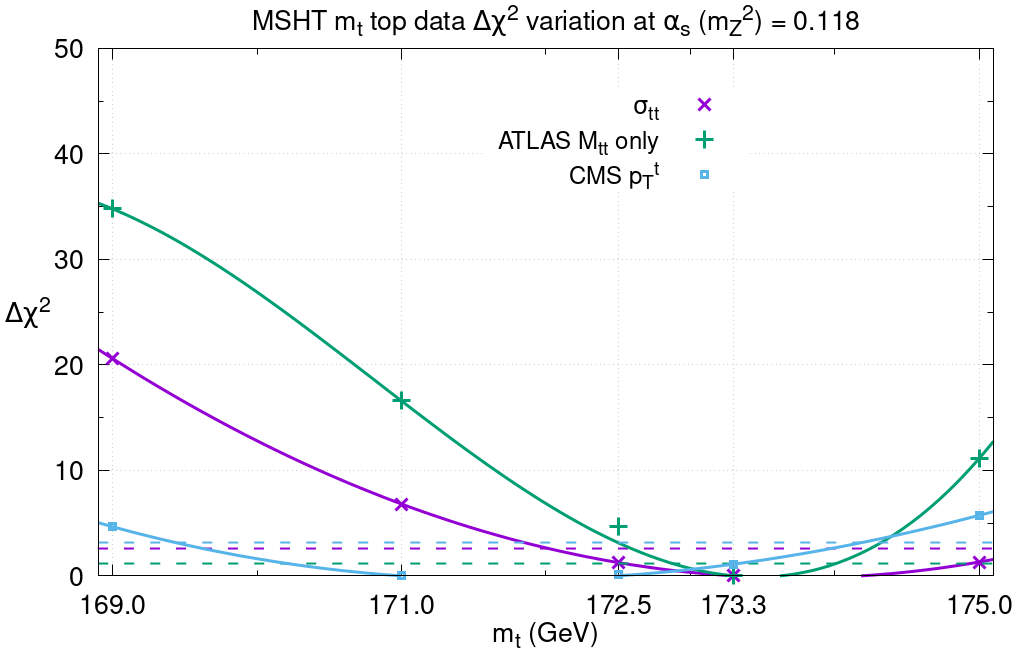}
\includegraphics[width=0.49\textwidth]{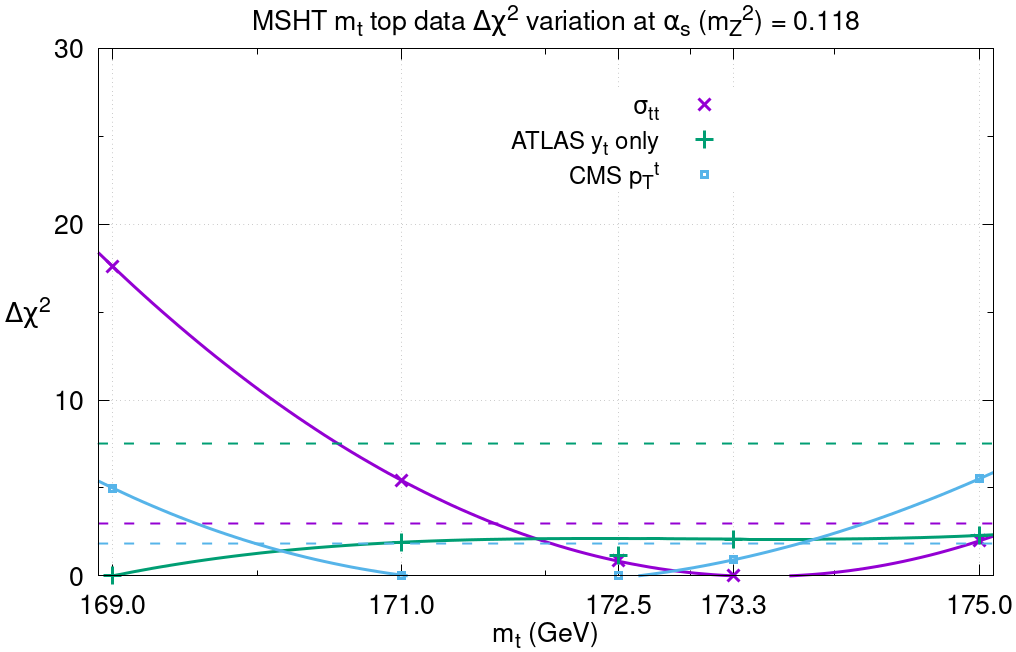}
\caption{$\Delta\chi^2$ of a single ATLAS dataset, the CMS $\pT$ data and the total cross section data as a function of $\mt$ and with fixed $\as=0.118$. Left: ATLAS $\mtt$. Right: ATLAS $\yt$.}
\label{ATLASsingle}
\end{figure}

We turn to consider including single ATLAS distributions, again refitting in every case. Having observed in sec.~\ref{CMSdistribution} that CMS rapidity distributions show a reduced sensitivity to the top-quark mass relative to the $\pT$ and $\mtt$ distributions (as expected theoretically), and verified this for the pairs of ATLAS distributions, we wish to gauge the extent to which this is the case for the individual ATLAS distributions. To that end, we repeat our fits using single ATLAS distributions, choosing either the $\mtt$ or $\yt$. This also allows us to assess the effect of the ATLAS data in a `clean' environment, without making any assumptions about the correlations (or otherwise) between distributions. We present our results in fig.~\ref{ATLASsingle}.

We note that by including only the ATLAS $\mtt$ distribution we retain significant constraints on the top-quark mass -- this single kinematic variable is sufficient to provide relatively tight bounds on $\mt$ (the same holds for $\pT)$. Using only the ATLAS $\yt$ (or indeed $\ytt$) instead has a very large and detrimental effect on the sensitivity, and all constraining power of this dataset is effectively lost. This confirms our na\"ive expectation and also verifies that our ability to place constraints on $\mt$ using the ATLAS data is largely independent of the exact correlation model between distributions.

\section{Study of $\alpha_s$ sensitivity}\label{alphas}
An advantage of using differential top-quark data is their ability to constrain both the top-quark mass and the strong coupling. Although extractions of the strong coupling using pair production cross section data alone have been performed~\cite{CMS:2014rml,Klijnsma:2017eqp,CMS:2018fks,ATLAS:2019hau}, only by supplementing this with differential information is it possible to extract also the top-quark mass~\cite{CMS:2019esx,Cooper-Sarkar:2020twv}. Our focus in this work so far has been on the top-quark mass, largely because the MSHT global PDF fit already contains several datasets able to place stringent bounds on $\as$ \cite{Cridge:2021qfd} but which are less able to bound $\mt$, as seen in fig.~\ref{heatmapnotopalltop} (right). In addition, the top-quark datasets also show somewhat greater sensitivity to $\mt$ than $\as$, see fig.~\ref{heatmapnotopalltop} (left). Nonetheless it is instructive to analyse their $\as$ sensitivities and the extent to which they are able to provide bounds competitive with other datasets in the global fit.

First, we analyse which datasets provide the tightest bounds on $\as$, both in the default setup of our analysis and also in the original MSHT20 $\as$ extraction. We remind the reader that these are expected to differ slightly, as a result of the exclusion of the ATLAS and CMS dilepton data and other minor changes. In the MSHT20 analysis of ref.~\cite{Cridge:2021qfd} the top-quark mass dependence was not available for the differential top-quark datasets, and so while $\as$ bounds were analysed, they were not used, given the potential for correlation between $\mt$ and the extracted $\as$ value. In this analysis, we now include the top-quark mass dependence for the single differential lepton+jets channel, which gives us the confidence to perform this study. We consider the $\as$ sensitivity in the $\mt=173.3~\mathrm{GeV}$ slice (given the results of sec.~\ref{sensitivity}) which contains the best overall global fit in the two-dimensional $\as-\mt$ plane and is closest to our best extracted mass value $\mt=173.0~\gev$. We present our findings in tab.~\ref{asbounds}.
 We have nonetheless verified that the results are similar at $\mt=172.5~\mathrm{GeV}$ -- we provide these results for the interested reader in appendix~\ref{appastable}. 
 
In the MSHT20 analysis the best fit at NNLO in QCD was found to be $\as=0.1174$, with the CMS 8~TeV $W$ data~\cite{CMS:2016qqr} providing a bound 0.0012 lower and the BCDMS proton data providing a bound 0.0013 higher, as indicated in Table~\ref{asbounds}. It was observed that for the top-quark data, the $\ttbar$ total cross section  data~\cite{ATLAS:2010zaw,ATLAS:2011whm,ATLAS:2012ptu,ATLAS:2012aa,ATLAS:2012xru,ATLAS:2012qtn,ATLAS:2015xtk,CMS:2012ahl,CMS:2012exf,CMS:2012hcu,CMS:2013nie,CMS:2013yjt,CMS:2013hon,CMS:2014btv,CMS:2015auz,CDF:2013hmv} was able to provide reasonable lower and upper bounds, albeit not competitive with the best bounds across the global fit. The ATLAS 8~TeV multi-differential $\ttbar$ lepton+jet data~\cite{ATLAS:2015lsn} did not provide competitive bounds. In contrast, the CMS 8~TeV single differential $\ytt$~\cite{CMS:2015rld} provided an upper bound on $\as$ almost as strong as that provided by BCDMS proton data \cite{BCDMS}, which is known to favour lower $\as$ values. We stress again, however, that due to the missing $\mt$ dependence this was not considered in the final MSHT20 quoted values.

\begin{figure}[ht!]
\centering
\includegraphics[width=0.9\textwidth]{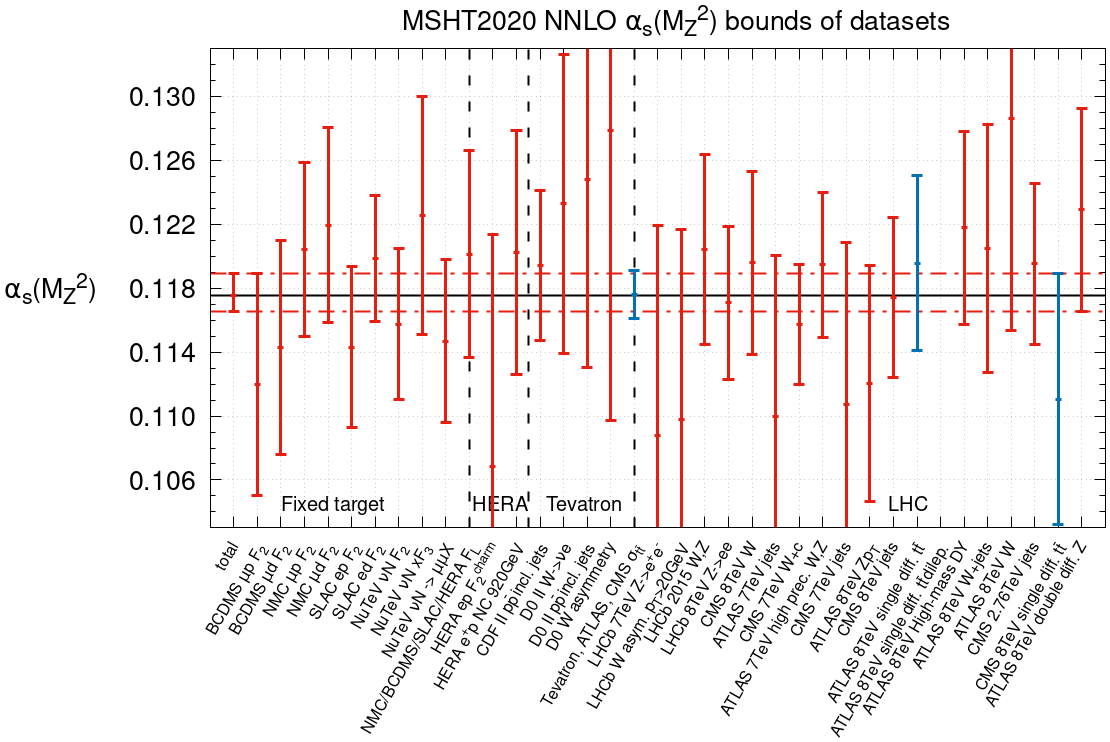}
\caption{$\as$ bounds placed by different datasets in the fits in this work at $\mt=173.3~{\rm GeV}$. This can be compared with fig.~6 (lower) from ref.~\cite{Cridge:2021qfd} or equivalently fig.~19 from ref.~\cite{dEnterria:2022hzv}. The total $t\bar{t}$ cross-section and single differential $t\bar{t}$ lepton+jet datasets now included the $m_t$ dependence and are shown in blue.}
\label{New_alphas_bounds_plot}
\end{figure}

\begin{figure}[ht!]
\centering
\includegraphics[width=0.75\textwidth]{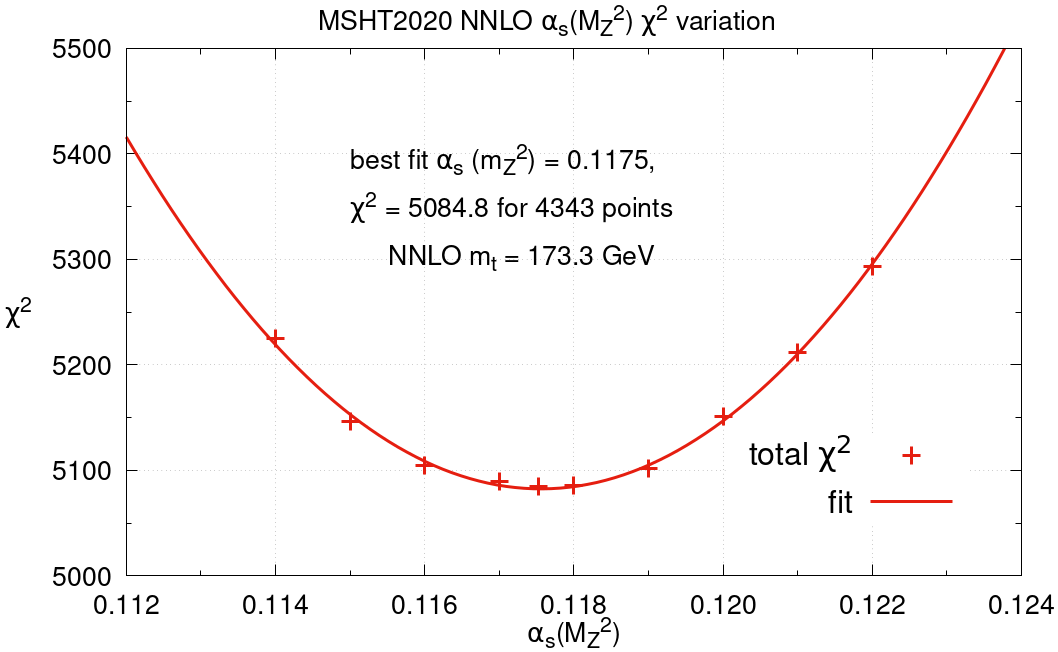}
\caption{The total $\chi^2$ of the whole global fit data as a function of $\as$, with $\mt=173.3~ {\rm GeV}$ taken for the top-quark datasets and at NNLO.}
\label{alphas_profile}
\end{figure}

We now find that the best fit at $\mt=173.3~\mathrm{GeV}$ is at $\as=0.1175$, in close agreement with ref. \cite{Cridge:2021qfd}. The most constraining lower bound of 0.1165 is provided by the ATLAS 8~TeV double differential $Z$ data~\cite{ATLAS:2017rue}. This dataset also provided a lower bound in the original MSHT20 fit, although not the most constraining. The CMS 8~TeV single-differential $\ytt$ in the lepton+jets channel now provides the strongest upper bound on $\as$ at 0.1189, essentially identical to the BCDMS proton data. The upper bound on $\as$ provided by the CMS 8~TeV $\ytt$ data reflects its preference for slightly lower theory predictions, as is also seen in its preference for large $\mt$ (i.e. it bounds $\mt$ from below) in sec.~\ref{topmassMSHT}. The total cross section data is again able to provide significant lower and upper bounds, in this fit now closer to the overall $\as$ bounds. Finally, the ATLAS $\ttbar$ data provide almost identical bounds to the MSHT20 analysis, and are equally poor in their constraining power. The bounds placed on $\as$ at NNLO by a selection of the most relevant datasets included in the MSHT20 global fit are shown in fig.~\ref{New_alphas_bounds_plot}, and exhibit good consistency with previous analyses in ref. \cite{Cridge:2021qfd}. In addition it demonstrates the competitive bounds placed by the top-quark datasets, shown in blue. The global fit bounds on $\as$ of -0.0010 and +0.0014 correspond to $\Delta\chi^2 = +10, +17$ respectively. The overall $\as$ $\chi^2$ profile is given in fig.~\ref{alphas_profile} at $\mt=173.3~{\rm GeV}$.

We note that the original MSHT20 analysis observed that the dileptonic $\ttbar$, single differential data were able to provide an upper bound on $\as$ similar to that given by the CMS 8~TeV $\ytt$. Since we do not have the theoretical predictions for these measurements for different top-quark masses, we do not consider this dataset in our analysis here (the corresponding entry is left vacant in fig.~\ref{New_alphas_bounds_plot}) and leave its examination to a future study. In addition, given the focus of this work is largely on the $\mt$ bounds, we have not here analysed the extent to which different CMS distributions for the single differential lepton+jet data may bound $\as$. We leave this to potential future work focused on $\as$ bounds in global PDF fits. 

\begin{table}[t!]
\begin{center}
\renewcommand\arraystretch{1.25}
\begin{threeparttable}
\begin{tabular}{|l|l|l|l|l|}
\hline
Data   & \multicolumn{2}{c|}{MSHT20}  &   \multicolumn{2}{c|}{This analysis}  \\
\hline
& Lower bound & Upper bound & Lower bound & Upper bound \\ 
\hline
All   &  0.1162 \tnote{$\ast$} & 0.1187 \tnote{$\dagger$} & 0.1165 \tnote{$\ddagger$} & 0.1189 \tnote{$\mathsection$}\\ 
$\sigma_{\ttbar}$   &  0.1144  & 0.1200  & 0.1161 & 0.1191 \\
CMS 8~TeV $\ytt$ &  0.1067  & 0.1187  & 0.1032 & 0.1189 \\
ATLAS $\ttbar$ &  0.1142  & 0.1255  & 0.1141 & 0.1251 \\
\hline
\end{tabular}
\begin{tablenotes}\footnotesize
\item[$\ast$] CMS 8~TeV $W$~\cite{CMS:2016qqr}\hspace{-1em}
\item[$\dagger$] BCDMS $p$~\cite{BCDMS}\hspace{-1em}
\item[$\ddagger$] ATLAS 8~TeV $Z~\rm{d. diff}$~\cite{ATLAS:2017rue}\hspace{-1em}
\item[$\mathsection$] CMS 8~TeV $\ytt$~\cite{CMS:2015rld}
\end{tablenotes}
\end{threeparttable}
\end{center}
\caption{Bounds on $\as$, obtained via a one-dimensional fit with $\mt=173.3~\gev$, for different datasets. In the case of the global fit where all datasets are considered, the single dataset giving rise to the tightest constraint is indicated. Results are shown for the original MSHT20 fit~\cite{Bailey:2020ooq} in ref.~\cite{Cridge:2021qfd} and for the fit we consider in this work, with dileptonic $\ttbar$ data removed. Results for $\mt=172.5~\gev$ appear in appendix~\ref{appastable}.}
\label{asbounds}   
\end{table}

\section{Impact on the gluon PDF}\label{PDF}
In this section, we examine the effect of different top-quark masses on the gluon parton distribution function. Top-quark pair production data is important in the context of global PDF fits, due to its ability to constrain the high-$x$ gluon PDF~\cite{Czakon:2013tha}. Indeed, a reduction in the high-$x$ gluon uncertainty motivated the choices of kinematic distribution from the ATLAS and CMS lepton+jets data for NNPDF3.1~\cite{Czakon:2016olj}. The small number of data points, can, however, limit the usefulness of these datasets in comparison to jet data for this purpose~\cite{Amat:2019upj,Hou:2019efy}. There is a general agreement among collaborations that the rapidity distributions provide the strongest constraints~\cite{Czakon:2016olj,Amat:2019upj,Hou:2019efy,Bailey:2019yze, Bailey:2020ooq}, although the ATLAS rapidity distributions seem to be in some tension with CMS jet data~\cite{Cridge:2021qfd}.

\begin{figure}[ht!]
\centering
\includegraphics[width=0.49\textwidth]{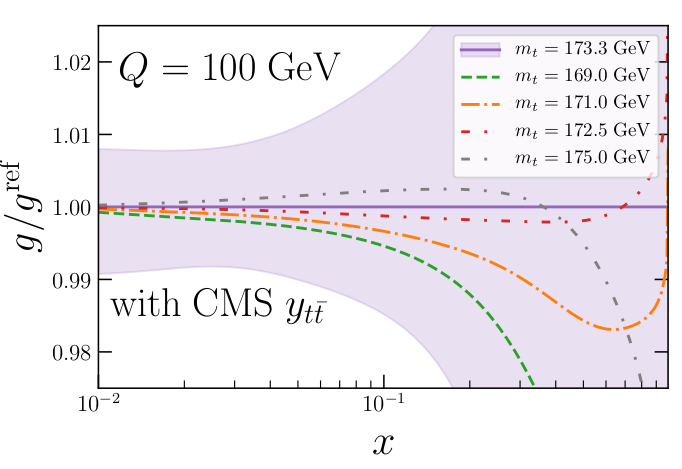}
\includegraphics[width=0.49\textwidth]{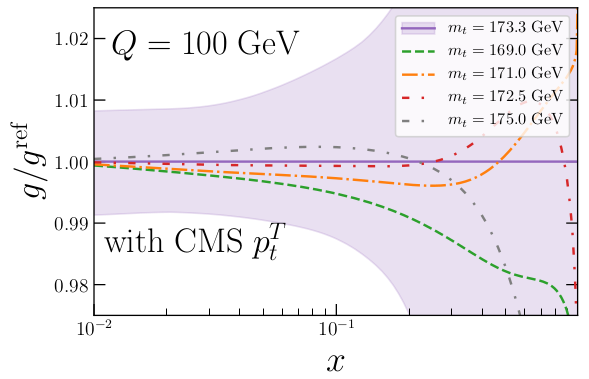}
\caption{Gluon PDF as a function of $x$ for various values of the top-quark mass $\mt$ and for $\as=0.118$. The ratio to the case $\mt=173.3~\gev$ is shown; PDF uncertainties are shown for this default $\mt$ value. Left: CMS $\ytt$ distribution. Right: CMS $\pT$ distribution.}
\label{gluon_mtdep}
\end{figure}

\begin{figure}[ht!]
\centering
\includegraphics[width=0.49\textwidth]{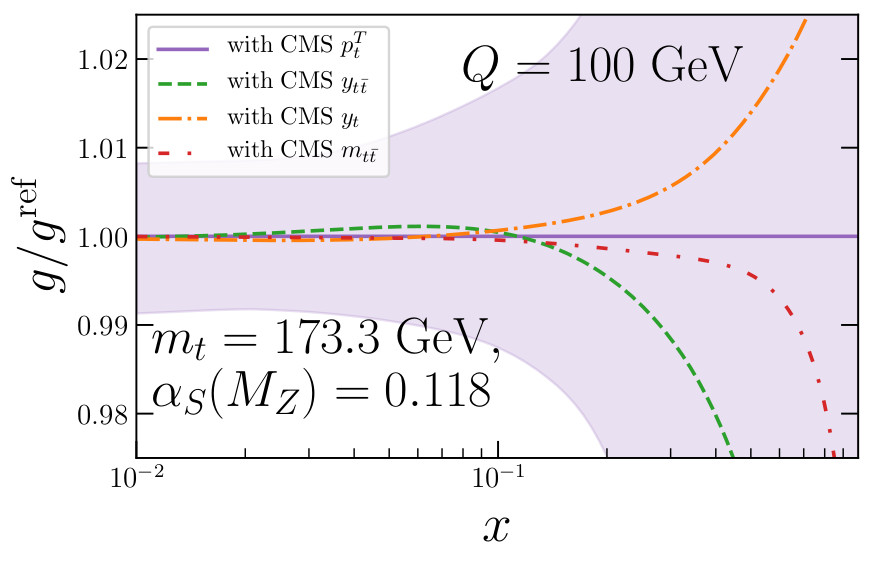}
\includegraphics[width=0.49\textwidth]{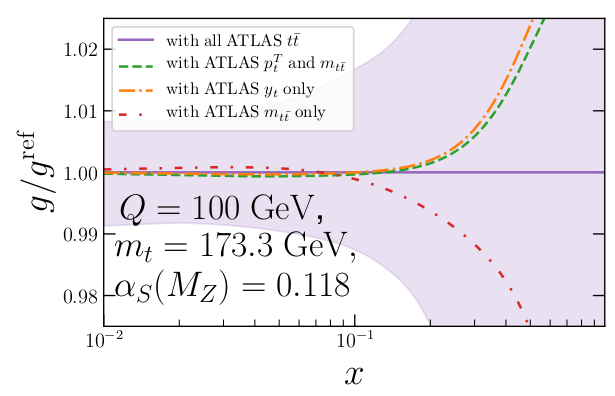}
\caption{Gluon PDF as a function of $x$ for various choices of included kinematic distribution. Fixed values of $\mt=173.3~\gev$ and $\as=0.118$ are considered. Left: all ATLAS and total cross section data included, with different choices for the CMS distribution. Right: all total cross section data and CMS $\pT$ distribution included, with various choices for the ATLAS distributions.}
\label{gluon_CMSandATLASaltdists}
\end{figure}

In fig.~\ref{gluon_mtdep} we show the gluon PDF $g(x)$ for various values of the top-quark mass. We show both the default MSHT20 choice using the CMS $\ytt$ distribution, as well as the alternative choice we investigate in sec.~\ref{distributions} of using the $\pT$ distribution. In both cases we normalise to the gluon PDF for $\mt=173.3~\gev$. We find the effect of changing $\mt$ is limited to large $x$ values and causes an increase of the PDF with increasing mass, as expected. Moreover, the effects are well within the PDF uncertainties, shown for the default value of $\mt=173.3~\gev$. Comparing the two choices of CMS distribution, we find a slightly greater dependence on the mass in the $\ytt$ case than in the $\pT$ case. We note that no data is available for $x \gtrsim 0.5$, and so beyond this range one has no clear physical interpretation. 

In fig.~\ref{gluon_CMSandATLASaltdists} we instead present the effects of changing the included CMS or ATLAS $t\bar{t}$ lepton + jet distributions on the gluon PDF relative to the baseline case of the CMS $\pT$ and all ATLAS distributions (with standard MSHT decorrelations described previously). The different choices of CMS distribution (left) have notable effects on the gluon at $x \gtrsim 0.1$, but are well within the large PDF uncertainties. Meanwhile the different choices of included ATLAS distributions (right) also lead to notable differences, reflecting both the changes of data included and the effects of decorrelations where multiple distributions are simultaneously included. This is consistent with ref.~\cite{Bailey:2020ooq} where it was observed that decorrelation in the ATLAS $t\bar{t}$ data has a significant effect on high $x$ gluon, albeit within the large PDF uncertainties in this region. 

\section{Summary of findings for $\mt$}\label{summary}
Our analysis in secs.~\ref{topmassMSHT} and~\ref{distributions} has shown that the single-differential ATLAS data in the lepton+jets channel are able to place relatively strong constraints on the top-quark mass. In contrast, the CMS data in the same channel generally provide weaker constraints, particularly when the $\ytt$ distribution is chosen (as is the case in the MSHT20 default setup), though the $\pT$ distribution provides the strongest constraint of the four choices available. With this in mind, we present our final results for $\mt$ using all available ATLAS distributions, the total cross section data and the CMS $\pT$ distribution (the bounds however are independent of the CMS distribution chosen, even after refitting). Maintaining a cubic parameterisation for the $\mt$ dependence, the global fit returns 
\begin{equation}
    \mt^{\rm pole}=173.0\pm0.6~\gev\,.
\end{equation}
The bounds in this case arise solely from the ATLAS data -- we note, however, that the lower bound obtained from the total cross section data is only $0.5~\gev$ lower, while the upper bound from the CMS $\pT$ distribution is only $0.4~\gev$ higher. We stress that the dynamic tolerance approach we adopt returns a conservative estimate of the uncertainty compared to a $\Delta\chi^2=1$ criterion (which would result in $\mt^{\rm pole}=173.0\pm0.3~\gev$). We also emphasise that our choice of a cubic interpolation provided the most conservative estimate of the uncertainty, fully containing the range of values obtained when using other reasonable interpolations.

Notably, the central value we obtain of $173.0~\gev$ is the same as seen in ref.~\cite{Kadir:2020yml} for the absolute data. We also remark that, considering the study on our $\as$ sensitivity in sec.~\ref{alphas}, our best fit value of $\mt$ lies between the values considered in tabs.~\ref{asbounds} ($\mt=173.3~\gev$) and~\ref{asboundsmt1725} ($\mt=172.5~\gev$). Given that the limits on $\as$ are very similar in those two cases, we might also expect the limits corresponding to our best fit value of $\mt$ to be comparable.\footnote{One should note however that the fits in sec.~\ref{alphas} use the CMS $\ytt$ distribution as in our default setup rather than the $\pT$ distribution we have adopted for the $\mt$ bounds. For both fixed values of $\mt=172.5~\gev$ and $\mt=173.3~\gev$, we find best fit values of $\as=0.1175$.} The Particle Data Group quote a pole mass from cross section measurements of $172.5\pm0.7~\gev$~\cite{Workman:2022ynf}. This is entirely compatible with our result, with a similar level of uncertainty. Furthermore, the authors of ref.~\cite{Gao:2022srd} find a central value of $\mt=172.58~\gev$ in the context of their global fit, which uses NNLO $K$-factors on NLO predictions to approximate the full NNLO mass dependence of the theory predictions. This is also consistent with our analysis.

\section{Conclusions}\label{conc}
In this work, we have examined the ability of differential measurements of the $\ttbar$ process to constrain the top-quark mass $\mt$. Specifically, we compared NNLO theory predictions for the top-quark transverse momentum $\pT$, the pair invariant mass $\mtt$ and the single and pair rapidities $\yt,\,\ytt$ with measurements taken by the ATLAS and CMS collaborations at a centre-of-mass energy of 8~TeV and in the lepton+jets channel. We performed our study in the context of the global MSHT20 PDF fit, thus fully accounting for any correlations between our parameters of interest and those of the parton distribution functions. 

We find that the combined ATLAS data provide stronger constraints on the top-quark mass than those from CMS. We have explored different choices of distribution and treatments of the experimental correlation matrices, finding that certain options, e.g. the ATLAS $\mtt$ distribution, are much better than others, e.g. rapidity distributions. This confirms theoretical expectations about the mass dependence of kinematic variables. In addition to the differential information, total cross section data are able to provide some constraining ability. 

We studied the sensitivity of the top-quark datasets to the strong coupling $\as$ and performed fits for a fixed value of $\mt=173.3~\gev$ using the same lepton+jets data as present in the original MSHT20 extraction. We found that the CMS $\ytt$ distribution in particular was able to provide an upper bound on $\as$, competitive with the BCDMS dataset which proved most tightly constraining in the original fit. While the total cross section data was also able to provide reasonable limits on $\as$, the ATLAS data was significantly less useful for this purpose.

Finally, we examined the effect of different top-quark masses on the gluon PDF. We observed that higher values of $\mt$ result in an upwards shift of the high-$x$ gluon, as expected, but remain well within PDF uncertainties. We also noted that for fixed values of the mass, the choice of distribution included in the fit had a significant impact, albeit within the large uncertainties.

In further studies, it would be interesting to examine the impact of including both the single differential ATLAS data taken in the dilepton channel~\cite{ATLAS:2016pal} as well as the double-differential CMS data~\cite{CMS:2017iqf}, since both of these datasets are included in the MSHT20 fit. Our ability to do so relies on the availability of NNLO computations for these distributions at different values of the top-quark mass. Similarly, inclusion of the 13~TeV datasets would be a natural extension. The difference in constraining power of absolute and normalised data could also be investigated, as could fits using theory calculations based on resummed predictions, see e.g. refs.~\cite{Li:2013mia,Catani:2018mei,Alioli:2021ggd,Ju:2022wia}. Additionally, it may be interesting to investigate the effect of different scale choices on the fit, again dependent on the availability of theoretical predictions. Finally, an examination of the impact of the different choices of distribution on the gluon uncertainty could be undertaken. We leave these topics to potential future work.

\backmatter

\bmhead{Acknowledgments}

We thank M.~Czakon and A.~Mitov for making the NNLO grids needed for this work available. We are indebted to R.~Thorne and L.~A.~Harland-Lang for discussions and comments on the manuscript. We also thank the Lawrence Berkeley National Laboratory for hospitality while this work was finalised. TC acknowledges that this project has received funding from the European Research Council (ERC) under the European Union’s Horizon 2020 research and innovation programme (Grant agreement No. 101002090 COLORFREE), and also from STFC via grant awards ST/P000274/1 and ST/T000856/1 in the initial stages of this work. This work has also been partially funded by the Deutsche Forschungsgemeinschaft (DFG, German Research Foundation) - 491245950. MAL is supported by the UKRI guarantee scheme for the Marie Sk\l{}odowska-Curie postdoctoral fellowship, grant ref. EP/X021416/1. 

\begin{appendices}

\section{Strong coupling results for $\mt=172.5~\gev$} \label{appastable}

In this appendix we present analogous results to those in table~\ref{asbounds} of sec.~\ref{alphas}, but for a top-quark mass of $\mt = 172.5~\mathrm{GeV}$ - see table~\ref{asboundsmt1725}. We note that the ATLAS $8~\rm{TeV}$ double-differential data again provides an almost identical lower bound, while the $\ttbar$ total cross section and CMS $\ytt$ again provide competitive upper bounds. In this case, the $\ttbar$ total cross section data upper bound is now slightly more stringent and provides the global upper bound on $\as$.

\begin{table}[hb!]
\begin{center}
\renewcommand\arraystretch{1.25}
\begin{threeparttable}
\begin{tabular}{|l|l|l|l|l|}
\hline
Data   & \multicolumn{2}{c|}{MSHT20}  &   \multicolumn{2}{c|}{This analysis}  \\
\hline
& Lower bound & Upper bound & Lower bound & Upper bound \\ 
\hline
All   &  0.1162 \tnote{$\ast$} & 0.1187 \tnote{$\dagger$} & 0.1164 \tnote{$\ddagger$} & 0.1187 \tnote{$\mathsection$}\\ 
$\sigma_{\ttbar}$   &  0.1144  & 0.1200  & 0.1158 & 0.1187 \\
CMS 8~TeV $\ytt$ &  0.1067  & 0.1187  & 0.0999 & 0.1189 \\
ATLAS $\ttbar$ &  0.1142  & 0.1255  & No bound & No bound \\
\hline
\end{tabular}
\begin{tablenotes}\footnotesize
\item[$\ast$] CMS 8~TeV $W$~\cite{CMS:2016qqr}\hspace{-1em}
\item[$\dagger$] BCDMS $p$~\cite{BCDMS}\hspace{-1em}
\item[$\ddagger$] ATLAS 8~TeV $Z~\rm{d. diff}$~\cite{ATLAS:2017rue}\hspace{-1em}
\item[$\mathsection$] $\sigma_{\ttbar}$~\cite{ATLAS:2010zaw,ATLAS:2011whm,ATLAS:2012ptu,ATLAS:2012aa,ATLAS:2012xru,ATLAS:2012qtn,ATLAS:2015xtk,CMS:2012ahl,CMS:2012exf,CMS:2012hcu,CMS:2013nie,CMS:2013yjt,CMS:2013hon,CMS:2014btv,CMS:2015auz,CDF:2013hmv}
\end{tablenotes}
\end{threeparttable}
\end{center}
\caption{Bounds on $\as$, obtained via a one-dimensional fit with $\mt=172.5~\gev$, for different datasets. In the case of the global fit where all datasets are considered, the single dataset giving rise to the tightest constraint is indicated. 
Results are shown for the original MSHT20 fit in ref.~\cite{Bailey:2020ooq} and for the fit we consider in this work, with dileptonic $\ttbar$ data removed.
}
\label{asboundsmt1725}   
\end{table}

\end{appendices}
\newpage
\bibliography{MSHT_Top_Mass}


\begin{thebibliography}{97}
\ifx \bisbn   \undefined \def \bisbn  #1{ISBN #1}\fi
\ifx \binits  \undefined \def \binits#1{#1}\fi
\ifx \bauthor  \undefined \def \bauthor#1{#1}\fi
\ifx \batitle  \undefined \def \batitle#1{#1}\fi
\ifx \bjtitle  \undefined \def \bjtitle#1{#1}\fi
\ifx \bvolume  \undefined \def \bvolume#1{\textbf{#1}}\fi
\ifx \byear  \undefined \def \byear#1{#1}\fi
\ifx \bissue  \undefined \def \bissue#1{#1}\fi
\ifx \bfpage  \undefined \def \bfpage#1{#1}\fi
\ifx \blpage  \undefined \def \blpage #1{#1}\fi
\ifx \burl  \undefined \def \burl#1{\textsf{#1}}\fi
\ifx \doiurl  \undefined \def \doiurl#1{\url{https://doi.org/#1}}\fi
\ifx \betal  \undefined \def \betal{\textit{et al.}}\fi
\ifx \binstitute  \undefined \def \binstitute#1{#1}\fi
\ifx \binstitutionaled  \undefined \def \binstitutionaled#1{#1}\fi
\ifx \bctitle  \undefined \def \bctitle#1{#1}\fi
\ifx \beditor  \undefined \def \beditor#1{#1}\fi
\ifx \bpublisher  \undefined \def \bpublisher#1{#1}\fi
\ifx \bbtitle  \undefined \def \bbtitle#1{#1}\fi
\ifx \bedition  \undefined \def \bedition#1{#1}\fi
\ifx \bseriesno  \undefined \def \bseriesno#1{#1}\fi
\ifx \blocation  \undefined \def \blocation#1{#1}\fi
\ifx \bsertitle  \undefined \def \bsertitle#1{#1}\fi
\ifx \bsnm \undefined \def \bsnm#1{#1}\fi
\ifx \bsuffix \undefined \def \bsuffix#1{#1}\fi
\ifx \bparticle \undefined \def \bparticle#1{#1}\fi
\ifx \barticle \undefined \def \barticle#1{#1}\fi
\bibcommenthead
\ifx \bconfdate \undefined \def \bconfdate #1{#1}\fi
\ifx \botherref \undefined \def \botherref #1{#1}\fi
\ifx \url \undefined \def \url#1{\textsf{#1}}\fi
\ifx \bchapter \undefined \def \bchapter#1{#1}\fi
\ifx \bbook \undefined \def \bbook#1{#1}\fi
\ifx \bcomment \undefined \def \bcomment#1{#1}\fi
\ifx \oauthor \undefined \def \oauthor#1{#1}\fi
\ifx \citeauthoryear \undefined \def \citeauthoryear#1{#1}\fi
\ifx \endbibitem  \undefined \def \endbibitem {}\fi
\ifx \bconflocation  \undefined \def \bconflocation#1{#1}\fi
\ifx \arxivurl  \undefined \def \arxivurl#1{\textsf{#1}}\fi
\csname PreBibitemsHook\endcsname

\bibitem{Corcella:2019tgt}
\begin{barticle}
\bauthor{\bsnm{Corcella}, \binits{G.}}:
\batitle{{The top-quark mass: challenges in definition and determination}}.
\bjtitle{Front. in Phys.}
\bvolume{7},
\bfpage{54}
(\byear{2019})
{\href{https://arxiv.org/abs/1903.06574}{{arXiv:1903.06574}}}
{[hep-ph]}.
\doiurl{10.3389/fphy.2019.00054}
\end{barticle}
\endbibitem

\bibitem{Forte:2020pyp}
\begin{barticle}
\bauthor{\bsnm{Forte}, \binits{S.}},
\bauthor{\bsnm{Kassabov}, \binits{Z.}}:
\batitle{{Why $\alpha _s$ cannot be determined from hadronic processes without
  simultaneously determining the parton distributions}}.
\bjtitle{Eur. Phys. J. C}
\bvolume{80}(\bissue{3}),
\bfpage{182}
(\byear{2020})
{\href{https://arxiv.org/abs/2001.04986}{{arXiv:2001.04986}}}
{[hep-ph]}.
\doiurl{10.1140/epjc/s10052-020-7748-6}
\end{barticle}
\endbibitem

\bibitem{Ball:2018iqk}
\begin{barticle}
\bauthor{\bsnm{Ball}, \binits{R.D.}},
\bauthor{\bsnm{Carrazza}, \binits{S.}},
\bauthor{\bsnm{Del~Debbio}, \binits{L.}},
\bauthor{\bsnm{Forte}, \binits{S.}},
\bauthor{\bsnm{Kassabov}, \binits{Z.}},
\bauthor{\bsnm{Rojo}, \binits{J.}},
\bauthor{\bsnm{Slade}, \binits{E.}},
\bauthor{\bsnm{Ubiali}, \binits{M.}}:
\batitle{{Precision determination of the strong coupling constant within a
  global PDF analysis}}.
\bjtitle{Eur. Phys. J. C}
\bvolume{78}(\bissue{5}),
\bfpage{408}
(\byear{2018})
{\href{https://arxiv.org/abs/1802.03398}{{arXiv:1802.03398}}}
{[hep-ph]}.
\doiurl{10.1140/epjc/s10052-018-5897-7}
\end{barticle}
\endbibitem

\bibitem{Hou:2019efy}
\begin{barticle}
\bauthor{\bsnm{Hou}, \binits{T.-J.}}, \betal:
\batitle{{New CTEQ global analysis of quantum chromodynamics with
  high-precision data from the LHC}}.
\bjtitle{Phys. Rev. D}
\bvolume{103}(\bissue{1}),
\bfpage{014013}
(\byear{2021})
{\href{https://arxiv.org/abs/1912.10053}{{arXiv:1912.10053}}}
{[hep-ph]}.
\doiurl{10.1103/PhysRevD.103.014013}
\end{barticle}
\endbibitem

\bibitem{Cridge:2021qfd}
\begin{barticle}
\bauthor{\bsnm{Cridge}, \binits{T.}},
\bauthor{\bsnm{Harland-Lang}, \binits{L.A.}},
\bauthor{\bsnm{Martin}, \binits{A.D.}},
\bauthor{\bsnm{Thorne}, \binits{R.S.}}:
\batitle{{An investigation of the $\alpha _S$ and heavy quark mass dependence
  in the MSHT20 global PDF analysis}}.
\bjtitle{Eur. Phys. J. C}
\bvolume{81}(\bissue{8}),
\bfpage{744}
(\byear{2021})
{\href{https://arxiv.org/abs/2106.10289}{{arXiv:2106.10289}}}
{[hep-ph]}.
\doiurl{10.1140/epjc/s10052-021-09533-7}
\end{barticle}
\endbibitem

\bibitem{H1:2021xxi}
\begin{barticle}
\bauthor{\bsnm{Abt}, \binits{I.}}, \betal:
\batitle{{Impact of jet-production data on the next-to-next-to-leading-order
  determination of HERAPDF2.0 parton distributions}}.
\bjtitle{Eur. Phys. J. C}
\bvolume{82}(\bissue{3}),
\bfpage{243}
(\byear{2022})
{\href{https://arxiv.org/abs/2112.01120}{{arXiv:2112.01120}}}
{[hep-ex]}.
\doiurl{10.1140/epjc/s10052-022-10083-9}
\end{barticle}
\endbibitem

\bibitem{dEnterria:2022hzv}
\begin{botherref}
\oauthor{\bsnm{d'Enterria}, \binits{D.}}, et al.:
{The strong coupling constant: State of the art and the decade ahead}
(2022)
{\href{https://arxiv.org/abs/2203.08271}{{arXiv:2203.08271}}}
{[hep-ph]}
\end{botherref}
\endbibitem

\bibitem{Carrazza:2019sec}
\begin{barticle}
\bauthor{\bsnm{Carrazza}, \binits{S.}},
\bauthor{\bsnm{Degrande}, \binits{C.}},
\bauthor{\bsnm{Iranipour}, \binits{S.}},
\bauthor{\bsnm{Rojo}, \binits{J.}},
\bauthor{\bsnm{Ubiali}, \binits{M.}}:
\batitle{{Can New Physics hide inside the proton?}}
\bjtitle{Phys. Rev. Lett.}
\bvolume{123}(\bissue{13}),
\bfpage{132001}
(\byear{2019})
{\href{https://arxiv.org/abs/1905.05215}{{arXiv:1905.05215}}}
{[hep-ph]}.
\doiurl{10.1103/PhysRevLett.123.132001}
\end{barticle}
\endbibitem

\bibitem{Greljo:2021kvv}
\begin{barticle}
\bauthor{\bsnm{Greljo}, \binits{A.}},
\bauthor{\bsnm{Iranipour}, \binits{S.}},
\bauthor{\bsnm{Kassabov}, \binits{Z.}},
\bauthor{\bsnm{Madigan}, \binits{M.}},
\bauthor{\bsnm{Moore}, \binits{J.}},
\bauthor{\bsnm{Rojo}, \binits{J.}},
\bauthor{\bsnm{Ubiali}, \binits{M.}},
\bauthor{\bsnm{Voisey}, \binits{C.}}:
\batitle{{Parton distributions in the SMEFT from high-energy Drell-Yan tails}}.
\bjtitle{JHEP}
\bvolume{07},
\bfpage{122}
(\byear{2021})
{\href{https://arxiv.org/abs/2104.02723}{{arXiv:2104.02723}}}
{[hep-ph]}.
\doiurl{10.1007/JHEP07(2021)122}
\end{barticle}
\endbibitem

\bibitem{ATLAS:2014nxi}
\begin{barticle}
\bauthor{\bsnm{Aad}, \binits{G.}}, \betal:
\batitle{{Measurement of the $t\bar{t}$ production cross-section using $e\mu $
  events with b-tagged jets in pp collisions at $\sqrt{s}$ = 7 and 8
  $\,\mathrm{TeV}$ with the ATLAS detector}}.
\bjtitle{Eur. Phys. J. C}
\bvolume{74}(\bissue{10}),
\bfpage{3109}
(\byear{2014})
{\href{https://arxiv.org/abs/1406.5375}{{arXiv:1406.5375}}}
{[hep-ex]}.
\doiurl{10.1140/epjc/s10052-016-4501-2}.
\bcomment{[Addendum: Eur.Phys.J.C 76, 642 (2016)]}
\end{barticle}
\endbibitem

\bibitem{CMS:2016yys}
\begin{barticle}
\bauthor{\bsnm{Khachatryan}, \binits{V.}}, \betal:
\batitle{{Measurement of the t-tbar production cross section in the e-mu
  channel in proton-proton collisions at sqrt(s) = 7 and 8 TeV}}.
\bjtitle{JHEP}
\bvolume{08},
\bfpage{029}
(\byear{2016})
{\href{https://arxiv.org/abs/1603.02303}{{arXiv:1603.02303}}}
{[hep-ex]}.
\doiurl{10.1007/JHEP08(2016)029}
\end{barticle}
\endbibitem

\bibitem{CMS:2018fks}
\begin{barticle}
\bauthor{\bsnm{Sirunyan}, \binits{A.M.}}, \betal:
\batitle{{Measurement of the $\mathrm{t}\overline{\mathrm{t}}$ production cross
  section, the top quark mass, and the strong coupling constant using dilepton
  events in pp collisions at $\sqrt{s} =$ 13 TeV}}.
\bjtitle{Eur. Phys. J. C}
\bvolume{79}(\bissue{5}),
\bfpage{368}
(\byear{2019})
{\href{https://arxiv.org/abs/1812.10505}{{arXiv:1812.10505}}}
{[hep-ex]}.
\doiurl{10.1140/epjc/s10052-019-6863-8}
\end{barticle}
\endbibitem

\bibitem{Czakon:2011xx}
\begin{barticle}
\bauthor{\bsnm{Czakon}, \binits{M.}},
\bauthor{\bsnm{Mitov}, \binits{A.}}:
\batitle{{Top++: A Program for the Calculation of the Top-Pair Cross-Section at
  Hadron Colliders}}.
\bjtitle{Comput. Phys. Commun.}
\bvolume{185},
\bfpage{2930}
(\byear{2014})
{\href{https://arxiv.org/abs/1112.5675}{{arXiv:1112.5675}}}
{[hep-ph]}.
\doiurl{10.1016/j.cpc.2014.06.021}
\end{barticle}
\endbibitem

\bibitem{Czakon:2013goa}
\begin{barticle}
\bauthor{\bsnm{Czakon}, \binits{M.}},
\bauthor{\bsnm{Fiedler}, \binits{P.}},
\bauthor{\bsnm{Mitov}, \binits{A.}}:
\batitle{{Total Top-Quark Pair-Production Cross Section at Hadron Colliders
  Through $O(\alpha^4_S)$}}.
\bjtitle{Phys. Rev. Lett.}
\bvolume{110},
\bfpage{252004}
(\byear{2013})
{\href{https://arxiv.org/abs/1303.6254}{{arXiv:1303.6254}}}
{[hep-ph]}.
\doiurl{10.1103/PhysRevLett.110.252004}
\end{barticle}
\endbibitem

\bibitem{CDF:2014upy}
\begin{botherref}
{Combination of CDF and D0 Results on the Mass of the Top Quark using up to 9.7
  fb$^{-1}$ at the Tevatron}
(2014)
{\href{https://arxiv.org/abs/1407.2682}{{arXiv:1407.2682}}}
{[hep-ex]}
\end{botherref}
\endbibitem

\bibitem{ATLAS:2018fwq}
\begin{barticle}
\bauthor{\bsnm{Aaboud}, \binits{M.}}, \betal:
\batitle{{Measurement of the top quark mass in the $t\bar{t}\rightarrow $
  lepton+jets channel from $\sqrt{s}=8$ TeV ATLAS data and combination with
  previous results}}.
\bjtitle{Eur. Phys. J. C}
\bvolume{79}(\bissue{4}),
\bfpage{290}
(\byear{2019})
{\href{https://arxiv.org/abs/1810.01772}{{arXiv:1810.01772}}}
{[hep-ex]}.
\doiurl{10.1140/epjc/s10052-019-6757-9}
\end{barticle}
\endbibitem

\bibitem{CMS:2018tye}
\begin{barticle}
\bauthor{\bsnm{Sirunyan}, \binits{A.M.}}, \betal:
\batitle{{Measurement of the top quark mass in the all-jets final state at
  $\sqrt{s} =$ 13 TeV and combination with the lepton+jets channel}}.
\bjtitle{Eur. Phys. J. C}
\bvolume{79}(\bissue{4}),
\bfpage{313}
(\byear{2019})
{\href{https://arxiv.org/abs/1812.10534}{{arXiv:1812.10534}}}
{[hep-ex]}.
\doiurl{10.1140/epjc/s10052-019-6788-2}
\end{barticle}
\endbibitem

\bibitem{Gombas:2022zbh}
\begin{bchapter}
\bauthor{\bsnm{Gombas}, \binits{J.}},
\bauthor{\bsnm{Fein}, \binits{J.}},
\bauthor{\bsnm{Sawford}, \binits{S.}},
\bauthor{\bsnm{Schwienhorst}, \binits{R.}}:
\bctitle{Dependence of the top-quark mass measured in top-quark pair production
  on the parton distribution functions at the lhc and future colliders}.
In: \bbtitle{2022 Snowmass Summer Study}
(\byear{2022})
\end{bchapter}
\endbibitem

\bibitem{ATLAS:2022jbw}
\begin{botherref}
{Measurement of the top-quark mass using a leptonic invariant mass in $pp$
  collisions at $\sqrt{s}=13~\textrm{TeV}$ with the ATLAS detector}
(2022)
{\href{https://arxiv.org/abs/2209.00583}{{arXiv:2209.00583}}}
{[hep-ex]}
\end{botherref}
\endbibitem

\bibitem{ATLAS:2022jpn}
\begin{botherref}
{Measurement of the top-quark mass in $t\bar{t} \to$ dilepton events with the
  ATLAS experiment using the template method in 13 TeV $pp$ collision data}
(2022)
\end{botherref}
\endbibitem

\bibitem{Guzzi:2014wia}
\begin{barticle}
\bauthor{\bsnm{Guzzi}, \binits{M.}},
\bauthor{\bsnm{Lipka}, \binits{K.}},
\bauthor{\bsnm{Moch}, \binits{S.-O.}}:
\batitle{{Top-quark pair production at hadron colliders: differential cross
  section and phenomenological applications with DiffTop}}.
\bjtitle{JHEP}
\bvolume{01},
\bfpage{082}
(\byear{2015})
{\href{https://arxiv.org/abs/1406.0386}{{arXiv:1406.0386}}}
{[hep-ph]}.
\doiurl{10.1007/JHEP01(2015)082}
\end{barticle}
\endbibitem

\bibitem{CMS:2019esx}
\begin{barticle}
\bauthor{\bsnm{Sirunyan}, \binits{A.M.}}, \betal:
\batitle{{Measurement of $\mathrm{t\bar t}$ normalised multi-differential cross
  sections in pp collisions at $\sqrt s=13$ TeV, and simultaneous determination
  of the strong coupling strength, top quark pole mass, and parton distribution
  functions}}.
\bjtitle{Eur. Phys. J. C}
\bvolume{80}(\bissue{7}),
\bfpage{658}
(\byear{2020})
{\href{https://arxiv.org/abs/1904.05237}{{arXiv:1904.05237}}}
{[hep-ex]}.
\doiurl{10.1140/epjc/s10052-020-7917-7}
\end{barticle}
\endbibitem

\bibitem{Kadir:2020yml}
\begin{barticle}
\bauthor{\bsnm{Kadir}, \binits{M.}},
\bauthor{\bsnm{Ablat}, \binits{A.}},
\bauthor{\bsnm{Dulat}, \binits{S.}},
\bauthor{\bsnm{Hou}, \binits{T.-J.}},
\bauthor{\bsnm{Sitiwaldi}, \binits{I.}}:
\batitle{{The impact of ATLAS and CMS single differential top-quark pair
  measurements at $\sqrt {s}$ = 8 TeV on CTEQ-TEA PDFs}}.
\bjtitle{Chin. Phys. C}
\bvolume{45}(\bissue{2}),
\bfpage{023111}
(\byear{2021})
{\href{https://arxiv.org/abs/2003.13740}{{arXiv:2003.13740}}}
{[hep-ph]}.
\doiurl{10.1088/1674-1137/abce10}
\end{barticle}
\endbibitem

\bibitem{Cooper-Sarkar:2020twv}
\begin{botherref}
\oauthor{\bsnm{Cooper-Sarkar}, \binits{A.M.}},
\oauthor{\bsnm{Czakon}, \binits{M.}},
\oauthor{\bsnm{Lim}, \binits{M.A.}},
\oauthor{\bsnm{Mitov}, \binits{A.}},
\oauthor{\bsnm{Papanastasiou}, \binits{A.S.}}:
{Simultaneous extraction of $\alpha_s$ and $m_t$ from LHC $t\bar{t}$
  differential distributions}
(2020)
{\href{https://arxiv.org/abs/2010.04171}{{arXiv:2010.04171}}}
{[hep-ph]}
\end{botherref}
\endbibitem

\bibitem{Gao:2022srd}
\begin{botherref}
\oauthor{\bsnm{Gao}, \binits{J.}},
\oauthor{\bsnm{Gao}, \binits{M.}},
\oauthor{\bsnm{Hobbs}, \binits{T.J.}},
\oauthor{\bsnm{Liu}, \binits{D.}},
\oauthor{\bsnm{Shen}, \binits{X.}}:
{Simultaneous CTEQ-TEA extraction of PDFs and SMEFT parameters from jet and
  $t{\bar t}$ data}
(2022)
{\href{https://arxiv.org/abs/2211.01094}{{arXiv:2211.01094}}}
{[hep-ph]}
\end{botherref}
\endbibitem

\bibitem{Alekhin:2016jjz}
\begin{barticle}
\bauthor{\bsnm{Alekhin}, \binits{S.}},
\bauthor{\bsnm{Moch}, \binits{S.}},
\bauthor{\bsnm{Thier}, \binits{S.}}:
\batitle{{Determination of the top-quark mass from hadro-production of single
  top-quarks}}.
\bjtitle{Phys. Lett. B}
\bvolume{763},
\bfpage{341}--\blpage{346}
(\byear{2016})
{\href{https://arxiv.org/abs/1608.05212}{{arXiv:1608.05212}}}
{[hep-ph]}.
\doiurl{10.1016/j.physletb.2016.10.062}
\end{barticle}
\endbibitem

\bibitem{CMS:2017mpr}
\begin{barticle}
\bauthor{\bsnm{Sirunyan}, \binits{A.M.}}, \betal:
\batitle{{Measurement of the top quark mass using single top quark events in
  proton-proton collisions at $\sqrt{s}= 8$ TeV}}.
\bjtitle{Eur. Phys. J. C}
\bvolume{77}(\bissue{5}),
\bfpage{354}
(\byear{2017})
{\href{https://arxiv.org/abs/1703.02530}{{arXiv:1703.02530}}}
{[hep-ex]}.
\doiurl{10.1140/epjc/s10052-017-4912-8}
\end{barticle}
\endbibitem

\bibitem{Gao:2020nhu}
\begin{barticle}
\bauthor{\bsnm{Gao}, \binits{M.S.}},
\bauthor{\bsnm{Yuan}, \binits{S.R.}},
\bauthor{\bsnm{Gao}, \binits{J.}}:
\batitle{{Top-quark mass determination from t-channel single top production at
  the LHC}}.
\bjtitle{JHEP}
\bvolume{04},
\bfpage{054}
(\byear{2021})
{\href{https://arxiv.org/abs/2007.15527}{{arXiv:2007.15527}}}
{[hep-ph]}.
\doiurl{10.1007/JHEP04(2021)054}
\end{barticle}
\endbibitem

\bibitem{Schwienhorst:2022yqu}
\begin{botherref}
\oauthor{\bsnm{Agashe}, \binits{K.}}, et al.:
{Report of the Topical Group on Top quark physics and heavy flavor production
  for Snowmass 2021}
(2022)
{\href{https://arxiv.org/abs/2209.11267}{{arXiv:2209.11267}}}
{[hep-ph]}
\end{botherref}
\endbibitem

\bibitem{ATLAS:2015lsn}
\begin{barticle}
\bauthor{\bsnm{Aad}, \binits{G.}}, \betal:
\batitle{{Measurements of top-quark pair differential cross-sections in the
  lepton+jets channel in $pp$ collisions at $\sqrt{s}=8$ TeV using the ATLAS
  detector}}.
\bjtitle{Eur. Phys. J. C}
\bvolume{76}(\bissue{10}),
\bfpage{538}
(\byear{2016})
{\href{https://arxiv.org/abs/1511.04716}{{arXiv:1511.04716}}}
{[hep-ex]}.
\doiurl{10.1140/epjc/s10052-016-4366-4}
\end{barticle}
\endbibitem

\bibitem{CMS:2015rld}
\begin{barticle}
\bauthor{\bsnm{Khachatryan}, \binits{V.}}, \betal:
\batitle{{Measurement of the differential cross section for top quark pair
  production in pp collisions at $\sqrt{s} = 8\,\text {TeV} $}}.
\bjtitle{Eur. Phys. J. C}
\bvolume{75}(\bissue{11}),
\bfpage{542}
(\byear{2015})
{\href{https://arxiv.org/abs/1505.04480}{{arXiv:1505.04480}}}
{[hep-ex]}.
\doiurl{10.1140/epjc/s10052-015-3709-x}
\end{barticle}
\endbibitem

\bibitem{Bailey:2020ooq}
\begin{barticle}
\bauthor{\bsnm{Bailey}, \binits{S.}},
\bauthor{\bsnm{Cridge}, \binits{T.}},
\bauthor{\bsnm{Harland-Lang}, \binits{L.A.}},
\bauthor{\bsnm{Martin}, \binits{A.D.}},
\bauthor{\bsnm{Thorne}, \binits{R.S.}}:
\batitle{{Parton distributions from LHC, HERA, Tevatron and fixed target data:
  MSHT20 PDFs}}.
\bjtitle{Eur. Phys. J. C}
\bvolume{81}(\bissue{4}),
\bfpage{341}
(\byear{2021})
{\href{https://arxiv.org/abs/2012.04684}{{arXiv:2012.04684}}}
{[hep-ph]}.
\doiurl{10.1140/epjc/s10052-021-09057-0}
\end{barticle}
\endbibitem

\bibitem{ATLAS:2018owm}
\begin{botherref}
{Determination of the parton distribution functions of the proton from ATLAS
  measurements of differential $W$ and $Z/\gamma^*$ and $t\bar{t}$ cross
  sections}
(2018)
\end{botherref}
\endbibitem

\bibitem{Czakon:2016olj}
\begin{barticle}
\bauthor{\bsnm{Czakon}, \binits{M.}},
\bauthor{\bsnm{Hartland}, \binits{N.P.}},
\bauthor{\bsnm{Mitov}, \binits{A.}},
\bauthor{\bsnm{Nocera}, \binits{E.R.}},
\bauthor{\bsnm{Rojo}, \binits{J.}}:
\batitle{{Pinning down the large-x gluon with NNLO top-quark pair differential
  distributions}}.
\bjtitle{JHEP}
\bvolume{04},
\bfpage{044}
(\byear{2017})
{\href{https://arxiv.org/abs/1611.08609}{{arXiv:1611.08609}}}
{[hep-ph]}.
\doiurl{10.1007/JHEP04(2017)044}
\end{barticle}
\endbibitem

\bibitem{Bailey:2019yze}
\begin{barticle}
\bauthor{\bsnm{Bailey}, \binits{S.}},
\bauthor{\bsnm{Harland-Lang}, \binits{L.}}:
\batitle{{Differential Top Quark Pair Production at the LHC: Challenges for PDF
  Fits}}.
\bjtitle{Eur. Phys. J. C}
\bvolume{80}(\bissue{1}),
\bfpage{60}
(\byear{2020})
{\href{https://arxiv.org/abs/1909.10541}{{arXiv:1909.10541}}}
{[hep-ph]}.
\doiurl{10.1140/epjc/s10052-020-7633-3}
\end{barticle}
\endbibitem

\bibitem{Thorne:2019mpt}
\begin{barticle}
\bauthor{\bsnm{Thorne}, \binits{R.S.}},
\bauthor{\bsnm{Bailey}, \binits{S.}},
\bauthor{\bsnm{Cridge}, \binits{T.}},
\bauthor{\bsnm{Harland-Lang}, \binits{L.A.}},
\bauthor{\bsnm{Martin}, \binits{A.D.}},
\bauthor{\bsnm{Nathvani}, \binits{R.}}:
\batitle{{Updates of PDFs using the MMHT framework}}.
\bjtitle{PoS}
\bvolume{DIS2019},
\bfpage{036}
(\byear{2019})
{\href{https://arxiv.org/abs/1907.08147}{{arXiv:1907.08147}}}
{[hep-ph]}.
\doiurl{10.22323/1.352.0036}
\end{barticle}
\endbibitem

\bibitem{Amat:2019upj}
\begin{botherref}
\oauthor{\bsnm{Amat}, \binits{O.}}, et al.:
{Impact of LHC top-quark pair measurements to CTEQ-TEA PDF analysis}
(2019)
{\href{https://arxiv.org/abs/1908.06441}{{arXiv:1908.06441}}}
{[hep-ph]}
\end{botherref}
\endbibitem

\bibitem{Hou:2019gfw}
\begin{barticle}
\bauthor{\bsnm{Hou}, \binits{T.-J.}},
\bauthor{\bsnm{Yu}, \binits{Z.}},
\bauthor{\bsnm{Dulat}, \binits{S.}},
\bauthor{\bsnm{Schmidt}, \binits{C.}},
\bauthor{\bsnm{Yuan}, \binits{C.-P.}}:
\batitle{{Updating and optimizing error parton distribution function sets in
  the Hessian approach. II.}}
\bjtitle{Phys. Rev. D}
\bvolume{100}(\bissue{11}),
\bfpage{114024}
(\byear{2019})
{\href{https://arxiv.org/abs/1907.12177}{{arXiv:1907.12177}}}
{[hep-ph]}.
\doiurl{10.1103/PhysRevD.100.114024}
\end{barticle}
\endbibitem

\bibitem{Amoroso:2020lgh}
\begin{bchapter}
\bauthor{\bsnm{Amoroso}, \binits{S.}}, \betal:
\bctitle{Les houches 2019: Physics at tev colliders: Standard model working
  group report}.
In: \bbtitle{11th Les Houches Workshop on Physics at TeV Colliders: PhysTeV Les
  Houches}
(\byear{2020})
\end{bchapter}
\endbibitem

\bibitem{NNPDF:2021njg}
\begin{barticle}
\bauthor{\bsnm{Ball}, \binits{R.D.}}, \betal:
\batitle{{The path to proton structure at 1\% accuracy}}.
\bjtitle{Eur. Phys. J. C}
\bvolume{82}(\bissue{5}),
\bfpage{428}
(\byear{2022})
{\href{https://arxiv.org/abs/2109.02653}{{arXiv:2109.02653}}}
{[hep-ph]}.
\doiurl{10.1140/epjc/s10052-022-10328-7}
\end{barticle}
\endbibitem

\bibitem{Amoroso:2022eow}
\begin{barticle}
\bauthor{\bsnm{Amoroso}, \binits{S.}}, \betal:
\batitle{{Snowmass 2021 Whitepaper: Proton Structure at the Precision
  Frontier}}.
\bjtitle{Acta Phys. Polon. B}
\bvolume{53}(\bissue{12}),
\bfpage{1}
(\byear{2022})
{\href{https://arxiv.org/abs/2203.13923}{{arXiv:2203.13923}}}
{[hep-ph]}.
\doiurl{10.5506/APhysPolB.53.12-A1}
\end{barticle}
\endbibitem

\bibitem{Harland-Lang:2014zoa}
\begin{barticle}
\bauthor{\bsnm{Harland-Lang}, \binits{L.A.}},
\bauthor{\bsnm{Martin}, \binits{A.D.}},
\bauthor{\bsnm{Motylinski}, \binits{P.}},
\bauthor{\bsnm{Thorne}, \binits{R.S.}}:
\batitle{{Parton distributions in the LHC era: MMHT 2014 PDFs}}.
\bjtitle{Eur. Phys. J. C}
\bvolume{75}(\bissue{5}),
\bfpage{204}
(\byear{2015})
{\href{https://arxiv.org/abs/1412.3989}{{arXiv:1412.3989}}}
{[hep-ph]}.
\doiurl{10.1140/epjc/s10052-015-3397-6}
\end{barticle}
\endbibitem

\bibitem{ATLAS:2016pal}
\begin{barticle}
\bauthor{\bsnm{Aaboud}, \binits{M.}}, \betal:
\batitle{{Measurement of top quark pair differential cross-sections in the
  dilepton channel in $pp$ collisions at $\sqrt{s}$ = 7 and 8 TeV with ATLAS}}.
\bjtitle{Phys. Rev. D}
\bvolume{94}(\bissue{9}),
\bfpage{092003}
(\byear{2016})
{\href{https://arxiv.org/abs/1607.07281}{{arXiv:1607.07281}}}
{[hep-ex]}.
\doiurl{10.1103/PhysRevD.94.092003}.
\bcomment{[Addendum: Phys.Rev.D 101, 119901 (2020)]}
\end{barticle}
\endbibitem

\bibitem{CMS:2017iqf}
\begin{barticle}
\bauthor{\bsnm{Sirunyan}, \binits{A.M.}}, \betal:
\batitle{{Measurement of double-differential cross sections for top quark pair
  production in pp collisions at $\sqrt{s} = 8$ $\,\text {TeV}$ and impact on
  parton distribution functions}}.
\bjtitle{Eur. Phys. J. C}
\bvolume{77}(\bissue{7}),
\bfpage{459}
(\byear{2017})
{\href{https://arxiv.org/abs/1703.01630}{{arXiv:1703.01630}}}
{[hep-ex]}.
\doiurl{10.1140/epjc/s10052-017-4984-5}
\end{barticle}
\endbibitem

\bibitem{ATLAS:2010zaw}
\begin{barticle}
\bauthor{\bsnm{Aad}, \binits{G.}}, \betal:
\batitle{{Measurement of the top quark-pair production cross section with ATLAS
  in pp collisions at $\sqrt{s}=7$ TeV}}.
\bjtitle{Eur. Phys. J. C}
\bvolume{71},
\bfpage{1577}
(\byear{2011})
{\href{https://arxiv.org/abs/1012.1792}{{arXiv:1012.1792}}}
{[hep-ex]}.
\doiurl{10.1140/epjc/s10052-011-1577-6}
\end{barticle}
\endbibitem

\bibitem{ATLAS:2011whm}
\begin{barticle}
\bauthor{\bsnm{Aad}, \binits{G.}}, \betal:
\batitle{{Measurement of the top quark pair production cross section in $pp$
  collisions at $\sqrt{s}=7$ TeV in dilepton final states with ATLAS}}.
\bjtitle{Phys. Lett. B}
\bvolume{707},
\bfpage{459}--\blpage{477}
(\byear{2012})
{\href{https://arxiv.org/abs/1108.3699}{{arXiv:1108.3699}}}
{[hep-ex]}.
\doiurl{10.1016/j.physletb.2011.12.055}
\end{barticle}
\endbibitem

\bibitem{ATLAS:2012ptu}
\begin{barticle}
\bauthor{\bsnm{Aad}, \binits{G.}}, \betal:
\batitle{{Measurement of the top quark pair production cross-section with ATLAS
  in the single lepton channel}}.
\bjtitle{Phys. Lett. B}
\bvolume{711},
\bfpage{244}--\blpage{263}
(\byear{2012})
{\href{https://arxiv.org/abs/1201.1889}{{arXiv:1201.1889}}}
{[hep-ex]}.
\doiurl{10.1016/j.physletb.2012.03.083}
\end{barticle}
\endbibitem

\bibitem{ATLAS:2012aa}
\begin{barticle}
\bauthor{\bsnm{Aad}, \binits{G.}}, \betal:
\batitle{{Measurement of the cross section for top-quark pair production in
  $pp$ collisions at $\sqrt{s}=7$ TeV with the ATLAS detector using final
  states with two high-pt leptons}}.
\bjtitle{JHEP}
\bvolume{05},
\bfpage{059}
(\byear{2012})
{\href{https://arxiv.org/abs/1202.4892}{{arXiv:1202.4892}}}
{[hep-ex]}.
\doiurl{10.1007/JHEP05(2012)059}
\end{barticle}
\endbibitem

\bibitem{ATLAS:2012xru}
\begin{barticle}
\bauthor{\bsnm{Aad}, \binits{G.}}, \betal:
\batitle{{Measurement of the top quark pair cross section with ATLAS in $pp$
  collisions at $\sqrt{s} =$ 7 TeV using final states with an electron or a
  muon and a hadronically decaying $\tau$ lepton}}.
\bjtitle{Phys. Lett. B}
\bvolume{717},
\bfpage{89}--\blpage{108}
(\byear{2012})
{\href{https://arxiv.org/abs/1205.2067}{{arXiv:1205.2067}}}
{[hep-ex]}.
\doiurl{10.1016/j.physletb.2012.09.032}
\end{barticle}
\endbibitem

\bibitem{ATLAS:2012qtn}
\begin{barticle}
\bauthor{\bsnm{Aad}, \binits{G.}}, \betal:
\batitle{{Measurement of the ttbar production cross section in the tau+jets
  channel using the ATLAS detector}}.
\bjtitle{Eur. Phys. J. C}
\bvolume{73}(\bissue{3}),
\bfpage{2328}
(\byear{2013})
{\href{https://arxiv.org/abs/1211.7205}{{arXiv:1211.7205}}}
{[hep-ex]}.
\doiurl{10.1140/epjc/s10052-013-2328-7}
\end{barticle}
\endbibitem

\bibitem{ATLAS:2015xtk}
\begin{barticle}
\bauthor{\bsnm{Aad}, \binits{G.}}, \betal:
\batitle{{Measurement of the top pair production cross section in 8 TeV
  proton-proton collisions using kinematic information in the lepton+jets final
  state with ATLAS}}.
\bjtitle{Phys. Rev. D}
\bvolume{91}(\bissue{11}),
\bfpage{112013}
(\byear{2015})
{\href{https://arxiv.org/abs/1504.04251}{{arXiv:1504.04251}}}
{[hep-ex]}.
\doiurl{10.1103/PhysRevD.91.112013}
\end{barticle}
\endbibitem

\bibitem{CMS:2012ahl}
\begin{barticle}
\bauthor{\bsnm{Chatrchyan}, \binits{S.}}, \betal:
\batitle{{Measurement of the top quark pair production cross section in $pp$
  collisions at $\sqrt{s} = 7$ TeV in dilepton final states containing a
  $\tau$}}.
\bjtitle{Phys. Rev. D}
\bvolume{85},
\bfpage{112007}
(\byear{2012})
{\href{https://arxiv.org/abs/1203.6810}{{arXiv:1203.6810}}}
{[hep-ex]}.
\doiurl{10.1103/PhysRevD.85.112007}
\end{barticle}
\endbibitem

\bibitem{CMS:2012exf}
\begin{barticle}
\bauthor{\bsnm{Chatrchyan}, \binits{S.}}, \betal:
\batitle{{Measurement of the $t\bar{t}$ Production Cross Section in the
  Dilepton Channel in $pp$ Collisions at $\sqrt{s}=7$ TeV}}.
\bjtitle{JHEP}
\bvolume{11},
\bfpage{067}
(\byear{2012})
{\href{https://arxiv.org/abs/1208.2671}{{arXiv:1208.2671}}}
{[hep-ex]}.
\doiurl{10.1007/JHEP11(2012)067}
\end{barticle}
\endbibitem

\bibitem{CMS:2012hcu}
\begin{barticle}
\bauthor{\bsnm{Chatrchyan}, \binits{S.}}, \betal:
\batitle{{Measurement of the $t\bar{t}$ Production Cross Section in $pp$
  Collisions at $\sqrt{s}=7$ TeV with Lepton + Jets Final States}}.
\bjtitle{Phys. Lett. B}
\bvolume{720},
\bfpage{83}--\blpage{104}
(\byear{2013})
{\href{https://arxiv.org/abs/1212.6682}{{arXiv:1212.6682}}}
{[hep-ex]}.
\doiurl{10.1016/j.physletb.2013.02.021}
\end{barticle}
\endbibitem

\bibitem{CMS:2013nie}
\begin{barticle}
\bauthor{\bsnm{Chatrchyan}, \binits{S.}}, \betal:
\batitle{{Measurement of the $t\bar{t}$ Production Cross Section in the $\tau$
  +Jets Channel in $pp$ Collisions at $\sqrt{s} = 7$ TeV}}.
\bjtitle{Eur. Phys. J. C}
\bvolume{73}(\bissue{4}),
\bfpage{2386}
(\byear{2013})
{\href{https://arxiv.org/abs/1301.5755}{{arXiv:1301.5755}}}
{[hep-ex]}.
\doiurl{10.1140/epjc/s10052-013-2386-x}
\end{barticle}
\endbibitem

\bibitem{CMS:2013yjt}
\begin{barticle}
\bauthor{\bsnm{Chatrchyan}, \binits{S.}}, \betal:
\batitle{{Measurement of the $t\bar{t}$ Production Cross Section in the All-Jet
  Final State in pp Collisions at $\sqrt{s}$ = 7 TeV}}.
\bjtitle{JHEP}
\bvolume{05},
\bfpage{065}
(\byear{2013})
{\href{https://arxiv.org/abs/1302.0508}{{arXiv:1302.0508}}}
{[hep-ex]}.
\doiurl{10.1007/JHEP05(2013)065}
\end{barticle}
\endbibitem

\bibitem{CMS:2013hon}
\begin{barticle}
\bauthor{\bsnm{Chatrchyan}, \binits{S.}}, \betal:
\batitle{{Measurement of the $t \bar{t}$ production cross section in the
  dilepton channel in pp collisions at $\sqrt{s}$ = 8 TeV}}.
\bjtitle{JHEP}
\bvolume{02},
\bfpage{024}
(\byear{2014})
{\href{https://arxiv.org/abs/1312.7582}{{arXiv:1312.7582}}}
{[hep-ex]}.
\doiurl{10.1007/JHEP02(2014)024}.
\bcomment{[Erratum: JHEP 02, 102 (2014)]}
\end{barticle}
\endbibitem

\bibitem{CMS:2014btv}
\begin{barticle}
\bauthor{\bsnm{Khachatryan}, \binits{V.}}, \betal:
\batitle{{Measurement of the $t \bar t$ Production Cross Section in $pp$
  Collisions at $\sqrt s = 8$ TeV in Dilepton Final States Containing One
  $\tau$ Lepton}}.
\bjtitle{Phys. Lett. B}
\bvolume{739},
\bfpage{23}--\blpage{43}
(\byear{2014})
{\href{https://arxiv.org/abs/1407.6643}{{arXiv:1407.6643}}}
{[hep-ex]}.
\doiurl{10.1016/j.physletb.2014.10.032}
\end{barticle}
\endbibitem

\bibitem{CMS:2015auz}
\begin{barticle}
\bauthor{\bsnm{Khachatryan}, \binits{V.}}, \betal:
\batitle{{Measurement of the $\mathrm{t}\overline{{\mathrm{t}}}$ production
  cross section in the all-jets final state in pp collisions at $\sqrt{s}=8$
  $\,\text {TeV}$}}.
\bjtitle{Eur. Phys. J. C}
\bvolume{76}(\bissue{3}),
\bfpage{128}
(\byear{2016})
{\href{https://arxiv.org/abs/1509.06076}{{arXiv:1509.06076}}}
{[hep-ex]}.
\doiurl{10.1140/epjc/s10052-016-3956-5}
\end{barticle}
\endbibitem

\bibitem{CDF:2013hmv}
\begin{barticle}
\bauthor{\bsnm{Aaltonen}, \binits{T.A.}}, \betal:
\batitle{{Combination of Measurements of the Top-Quark Pair Production Cross
  Section from the Tevatron Collider}}.
\bjtitle{Phys. Rev. D}
\bvolume{89}(\bissue{7}),
\bfpage{072001}
(\byear{2014})
{\href{https://arxiv.org/abs/1309.7570}{{arXiv:1309.7570}}}
{[hep-ex]}.
\doiurl{10.1103/PhysRevD.89.072001}
\end{barticle}
\endbibitem

\bibitem{Carli:2010rw}
\begin{barticle}
\bauthor{\bsnm{Carli}, \binits{T.}},
\bauthor{\bsnm{Clements}, \binits{D.}},
\bauthor{\bsnm{Cooper-Sarkar}, \binits{A.}},
\bauthor{\bsnm{Gwenlan}, \binits{C.}},
\bauthor{\bsnm{Salam}, \binits{G.P.}},
\bauthor{\bsnm{Siegert}, \binits{F.}},
\bauthor{\bsnm{Starovoitov}, \binits{P.}},
\bauthor{\bsnm{Sutton}, \binits{M.}}:
\batitle{{A posteriori inclusion of parton density functions in NLO QCD
  final-state calculations at hadron colliders: The APPLGRID Project}}.
\bjtitle{Eur. Phys. J. C}
\bvolume{66},
\bfpage{503}--\blpage{524}
(\byear{2010})
{\href{https://arxiv.org/abs/0911.2985}{{arXiv:0911.2985}}}
{[hep-ph]}.
\doiurl{10.1140/epjc/s10052-010-1255-0}
\end{barticle}
\endbibitem

\bibitem{Campbell:2012uf}
\begin{barticle}
\bauthor{\bsnm{Campbell}, \binits{J.M.}},
\bauthor{\bsnm{Ellis}, \binits{R.K.}}:
\batitle{{Top-Quark Processes at NLO in Production and Decay}}.
\bjtitle{J. Phys. G}
\bvolume{42}(\bissue{1}),
\bfpage{015005}
(\byear{2015})
{\href{https://arxiv.org/abs/1204.1513}{{arXiv:1204.1513}}}
{[hep-ph]}.
\doiurl{10.1088/0954-3899/42/1/015005}
\end{barticle}
\endbibitem

\bibitem{Barnreuther:2012wtj}
\begin{barticle}
\bauthor{\bsnm{B\"arnreuther}, \binits{P.}},
\bauthor{\bsnm{Czakon}, \binits{M.}},
\bauthor{\bsnm{Mitov}, \binits{A.}}:
\batitle{{Percent Level Precision Physics at the Tevatron: First Genuine NNLO
  QCD Corrections to $q \bar{q} \to t \bar{t} + X$}}.
\bjtitle{Phys. Rev. Lett.}
\bvolume{109},
\bfpage{132001}
(\byear{2012})
{\href{https://arxiv.org/abs/1204.5201}{{arXiv:1204.5201}}}
{[hep-ph]}.
\doiurl{10.1103/PhysRevLett.109.132001}
\end{barticle}
\endbibitem

\bibitem{Czakon:2012zr}
\begin{barticle}
\bauthor{\bsnm{Czakon}, \binits{M.}},
\bauthor{\bsnm{Mitov}, \binits{A.}}:
\batitle{{NNLO corrections to top-pair production at hadron colliders: the
  all-fermionic scattering channels}}.
\bjtitle{JHEP}
\bvolume{12},
\bfpage{054}
(\byear{2012})
{\href{https://arxiv.org/abs/1207.0236}{{arXiv:1207.0236}}}
{[hep-ph]}.
\doiurl{10.1007/JHEP12(2012)054}
\end{barticle}
\endbibitem

\bibitem{Czakon:2012pz}
\begin{barticle}
\bauthor{\bsnm{Czakon}, \binits{M.}},
\bauthor{\bsnm{Mitov}, \binits{A.}}:
\batitle{{NNLO corrections to top pair production at hadron colliders: the
  quark-gluon reaction}}.
\bjtitle{JHEP}
\bvolume{01},
\bfpage{080}
(\byear{2013})
{\href{https://arxiv.org/abs/1210.6832}{{arXiv:1210.6832}}}
{[hep-ph]}.
\doiurl{10.1007/JHEP01(2013)080}
\end{barticle}
\endbibitem

\bibitem{Catani:1996dj}
\begin{barticle}
\bauthor{\bsnm{Catani}, \binits{S.}},
\bauthor{\bsnm{Mangano}, \binits{M.L.}},
\bauthor{\bsnm{Nason}, \binits{P.}},
\bauthor{\bsnm{Trentadue}, \binits{L.}}:
\batitle{{The Top cross-section in hadronic collisions}}.
\bjtitle{Phys. Lett. B}
\bvolume{378},
\bfpage{329}--\blpage{336}
(\byear{1996})
{\href{https://arxiv.org/abs/hep-ph/9602208}{{arXiv:hep-ph/9602208}}}.
\doiurl{10.1016/0370-2693(96)00387-5}
\end{barticle}
\endbibitem

\bibitem{Cacciari:2011hy}
\begin{barticle}
\bauthor{\bsnm{Cacciari}, \binits{M.}},
\bauthor{\bsnm{Czakon}, \binits{M.}},
\bauthor{\bsnm{Mangano}, \binits{M.}},
\bauthor{\bsnm{Mitov}, \binits{A.}},
\bauthor{\bsnm{Nason}, \binits{P.}}:
\batitle{{Top-pair production at hadron colliders with next-to-next-to-leading
  logarithmic soft-gluon resummation}}.
\bjtitle{Phys. Lett. B}
\bvolume{710},
\bfpage{612}--\blpage{622}
(\byear{2012})
{\href{https://arxiv.org/abs/1111.5869}{{arXiv:1111.5869}}}
{[hep-ph]}.
\doiurl{10.1016/j.physletb.2012.03.013}
\end{barticle}
\endbibitem

\bibitem{Catani:2019iny}
\begin{barticle}
\bauthor{\bsnm{Catani}, \binits{S.}},
\bauthor{\bsnm{Devoto}, \binits{S.}},
\bauthor{\bsnm{Grazzini}, \binits{M.}},
\bauthor{\bsnm{Kallweit}, \binits{S.}},
\bauthor{\bsnm{Mazzitelli}, \binits{J.}},
\bauthor{\bsnm{Sargsyan}, \binits{H.}}:
\batitle{{Top-quark pair hadroproduction at next-to-next-to-leading order in
  QCD}}.
\bjtitle{Phys. Rev. D}
\bvolume{99}(\bissue{5}),
\bfpage{051501}
(\byear{2019})
{\href{https://arxiv.org/abs/1901.04005}{{arXiv:1901.04005}}}
{[hep-ph]}.
\doiurl{10.1103/PhysRevD.99.051501}
\end{barticle}
\endbibitem

\bibitem{Catani:2019hip}
\begin{barticle}
\bauthor{\bsnm{Catani}, \binits{S.}},
\bauthor{\bsnm{Devoto}, \binits{S.}},
\bauthor{\bsnm{Grazzini}, \binits{M.}},
\bauthor{\bsnm{Kallweit}, \binits{S.}},
\bauthor{\bsnm{Mazzitelli}, \binits{J.}}:
\batitle{{Top-quark pair production at the LHC: Fully differential QCD
  predictions at NNLO}}.
\bjtitle{JHEP}
\bvolume{07},
\bfpage{100}
(\byear{2019})
{\href{https://arxiv.org/abs/1906.06535}{{arXiv:1906.06535}}}
{[hep-ph]}.
\doiurl{10.1007/JHEP07(2019)100}
\end{barticle}
\endbibitem

\bibitem{Czakon:2017wor}
\begin{barticle}
\bauthor{\bsnm{Czakon}, \binits{M.}},
\bauthor{\bsnm{Heymes}, \binits{D.}},
\bauthor{\bsnm{Mitov}, \binits{A.}},
\bauthor{\bsnm{Pagani}, \binits{D.}},
\bauthor{\bsnm{Tsinikos}, \binits{I.}},
\bauthor{\bsnm{Zaro}, \binits{M.}}:
\batitle{{Top-pair production at the LHC through NNLO QCD and NLO EW}}.
\bjtitle{JHEP}
\bvolume{10},
\bfpage{186}
(\byear{2017})
{\href{https://arxiv.org/abs/1705.04105}{{arXiv:1705.04105}}}
{[hep-ph]}.
\doiurl{10.1007/JHEP10(2017)186}
\end{barticle}
\endbibitem

\bibitem{Czakon:2019txp}
\begin{barticle}
\bauthor{\bsnm{Czakon}, \binits{M.L.}}, \betal:
\batitle{{Top quark pair production at complete NLO accuracy with NNLO+NNLL'
  corrections in QCD}}.
\bjtitle{Chin. Phys. C}
\bvolume{44}(\bissue{8}),
\bfpage{083104}
(\byear{2020})
{\href{https://arxiv.org/abs/1901.08281}{{arXiv:1901.08281}}}
{[hep-ph]}.
\doiurl{10.1088/1674-1137/44/8/083104}
\end{barticle}
\endbibitem

\bibitem{Czakon:2015owf}
\begin{barticle}
\bauthor{\bsnm{Czakon}, \binits{M.}},
\bauthor{\bsnm{Heymes}, \binits{D.}},
\bauthor{\bsnm{Mitov}, \binits{A.}}:
\batitle{{High-precision differential predictions for top-quark pairs at the
  LHC}}.
\bjtitle{Phys. Rev. Lett.}
\bvolume{116}(\bissue{8}),
\bfpage{082003}
(\byear{2016})
{\href{https://arxiv.org/abs/1511.00549}{{arXiv:1511.00549}}}
{[hep-ph]}.
\doiurl{10.1103/PhysRevLett.116.082003}
\end{barticle}
\endbibitem

\bibitem{Czakon:2016ckf}
\begin{barticle}
\bauthor{\bsnm{Czakon}, \binits{M.}},
\bauthor{\bsnm{Fiedler}, \binits{P.}},
\bauthor{\bsnm{Heymes}, \binits{D.}},
\bauthor{\bsnm{Mitov}, \binits{A.}}:
\batitle{{NNLO QCD predictions for fully-differential top-quark pair production
  at the Tevatron}}.
\bjtitle{JHEP}
\bvolume{05},
\bfpage{034}
(\byear{2016})
{\href{https://arxiv.org/abs/1601.05375}{{arXiv:1601.05375}}}
{[hep-ph]}.
\doiurl{10.1007/JHEP05(2016)034}
\end{barticle}
\endbibitem

\bibitem{Czakon:2016dgf}
\begin{barticle}
\bauthor{\bsnm{Czakon}, \binits{M.}},
\bauthor{\bsnm{Heymes}, \binits{D.}},
\bauthor{\bsnm{Mitov}, \binits{A.}}:
\batitle{{Dynamical scales for multi-TeV top-pair production at the LHC}}.
\bjtitle{JHEP}
\bvolume{04},
\bfpage{071}
(\byear{2017})
{\href{https://arxiv.org/abs/1606.03350}{{arXiv:1606.03350}}}
{[hep-ph]}.
\doiurl{10.1007/JHEP04(2017)071}
\end{barticle}
\endbibitem

\bibitem{Czakon:2010td}
\begin{barticle}
\bauthor{\bsnm{Czakon}, \binits{M.}}:
\batitle{{A novel subtraction scheme for double-real radiation at NNLO}}.
\bjtitle{Phys. Lett. B}
\bvolume{693},
\bfpage{259}--\blpage{268}
(\byear{2010})
{\href{https://arxiv.org/abs/1005.0274}{{arXiv:1005.0274}}}
{[hep-ph]}.
\doiurl{10.1016/j.physletb.2010.08.036}
\end{barticle}
\endbibitem

\bibitem{Czakon:2014oma}
\begin{barticle}
\bauthor{\bsnm{Czakon}, \binits{M.}},
\bauthor{\bsnm{Heymes}, \binits{D.}}:
\batitle{{Four-dimensional formulation of the sector-improved residue
  subtraction scheme}}.
\bjtitle{Nucl. Phys. B}
\bvolume{890},
\bfpage{152}--\blpage{227}
(\byear{2014})
{\href{https://arxiv.org/abs/1408.2500}{{arXiv:1408.2500}}}
{[hep-ph]}.
\doiurl{10.1016/j.nuclphysb.2014.11.006}
\end{barticle}
\endbibitem

\bibitem{Kluge:2006xs}
\begin{bchapter}
\bauthor{\bsnm{Kluge}, \binits{T.}},
\bauthor{\bsnm{Rabbertz}, \binits{K.}},
\bauthor{\bsnm{Wobisch}, \binits{M.}}:
\bctitle{Fastnlo: Fast pqcd calculations for pdf fits}.
In: \bbtitle{14th International Workshop on Deep Inelastic Scattering},
pp. \bfpage{483}--\blpage{486}
(\byear{2006}).
\doiurl{10.1142/9789812706706_0110}
\end{bchapter}
\endbibitem

\bibitem{Britzger:2012bs}
\begin{bchapter}
\bauthor{\bsnm{Britzger}, \binits{D.}},
\bauthor{\bsnm{Rabbertz}, \binits{K.}},
\bauthor{\bsnm{Stober}, \binits{F.}},
\bauthor{\bsnm{Wobisch}, \binits{M.}}:
\bctitle{New features in version 2 of the fastnlo project}.
In: \bbtitle{20th International Workshop on Deep-Inelastic Scattering and
  Related Subjects},
pp. \bfpage{217}--\blpage{221}
(\byear{2012}).
\doiurl{10.3204/DESY-PROC-2012-02/165}
\end{bchapter}
\endbibitem

\bibitem{Czakon:2017dip}
\begin{botherref}
\oauthor{\bsnm{Czakon}, \binits{M.}},
\oauthor{\bsnm{Heymes}, \binits{D.}},
\oauthor{\bsnm{Mitov}, \binits{A.}}:
{fastNLO tables for NNLO top-quark pair differential distributions}
(2017)
{\href{https://arxiv.org/abs/1704.08551}{{arXiv:1704.08551}}}
{[hep-ph]}
\end{botherref}
\endbibitem

\bibitem{Martin:2009iq}
\begin{barticle}
\bauthor{\bsnm{Martin}, \binits{A.D.}},
\bauthor{\bsnm{Stirling}, \binits{W.J.}},
\bauthor{\bsnm{Thorne}, \binits{R.S.}},
\bauthor{\bsnm{Watt}, \binits{G.}}:
\batitle{{Parton distributions for the LHC}}.
\bjtitle{Eur. Phys. J. C}
\bvolume{63},
\bfpage{189}--\blpage{285}
(\byear{2009})
{\href{https://arxiv.org/abs/0901.0002}{{arXiv:0901.0002}}}
{[hep-ph]}.
\doiurl{10.1140/epjc/s10052-009-1072-5}
\end{barticle}
\endbibitem

\bibitem{Harland-Lang:2015nxa}
\begin{barticle}
\bauthor{\bsnm{Harland-Lang}, \binits{L.A.}},
\bauthor{\bsnm{Martin}, \binits{A.D.}},
\bauthor{\bsnm{Motylinski}, \binits{P.}},
\bauthor{\bsnm{Thorne}, \binits{R.S.}}:
\batitle{{Uncertainties on $\alpha _S$ in the MMHT2014 global PDF analysis and
  implications for SM predictions}}.
\bjtitle{Eur. Phys. J. C}
\bvolume{75}(\bissue{9}),
\bfpage{435}
(\byear{2015})
{\href{https://arxiv.org/abs/1506.05682}{{arXiv:1506.05682}}}
{[hep-ph]}.
\doiurl{10.1140/epjc/s10052-015-3630-3}
\end{barticle}
\endbibitem

\bibitem{Czakon:2016vfr}
\begin{barticle}
\bauthor{\bsnm{Czakon}, \binits{M.}},
\bauthor{\bsnm{Heymes}, \binits{D.}},
\bauthor{\bsnm{Mitov}, \binits{A.}}:
\batitle{{Bump hunting in LHC $t\bar t$ events}}.
\bjtitle{Phys. Rev. D}
\bvolume{94}(\bissue{11}),
\bfpage{114033}
(\byear{2016})
{\href{https://arxiv.org/abs/1608.00765}{{arXiv:1608.00765}}}
{[hep-ph]}.
\doiurl{10.1103/PhysRevD.94.114033}
\end{barticle}
\endbibitem

\bibitem{Ju:2019mqc}
\begin{barticle}
\bauthor{\bsnm{Ju}, \binits{W.-L.}},
\bauthor{\bsnm{Wang}, \binits{G.}},
\bauthor{\bsnm{Wang}, \binits{X.}},
\bauthor{\bsnm{Xu}, \binits{X.}},
\bauthor{\bsnm{Xu}, \binits{Y.}},
\bauthor{\bsnm{Yang}, \binits{L.L.}}:
\batitle{{Invariant-mass distribution of top-quark pairs and top-quark mass
  determination}}.
\bjtitle{Chin. Phys. C}
\bvolume{44}(\bissue{9}),
\bfpage{091001}
(\byear{2020})
{\href{https://arxiv.org/abs/1908.02179}{{arXiv:1908.02179}}}
{[hep-ph]}.
\doiurl{10.1088/1674-1137/44/9/091001}
\end{barticle}
\endbibitem

\bibitem{NNPDF:2017mvq}
\begin{barticle}
\bauthor{\bsnm{Ball}, \binits{R.D.}}, \betal:
\batitle{{Parton distributions from high-precision collider data}}.
\bjtitle{Eur. Phys. J. C}
\bvolume{77}(\bissue{10}),
\bfpage{663}
(\byear{2017})
{\href{https://arxiv.org/abs/1706.00428}{{arXiv:1706.00428}}}
{[hep-ph]}.
\doiurl{10.1140/epjc/s10052-017-5199-5}
\end{barticle}
\endbibitem

\bibitem{ATLAS:2017kux}
\begin{barticle}
\bauthor{\bsnm{Aaboud}, \binits{M.}}, \betal:
\batitle{{Measurement of the inclusive jet cross-sections in proton-proton
  collisions at $ \sqrt{s}=8 $ TeV with the ATLAS detector}}.
\bjtitle{JHEP}
\bvolume{09},
\bfpage{020}
(\byear{2017})
{\href{https://arxiv.org/abs/1706.03192}{{arXiv:1706.03192}}}
{[hep-ex]}.
\doiurl{10.1007/JHEP09(2017)020}
\end{barticle}
\endbibitem

\bibitem{CMS:2014rml}
\begin{barticle}
\bauthor{\bsnm{Chatrchyan}, \binits{S.}}, \betal:
\batitle{{Determination of the Top-Quark Pole Mass and Strong Coupling Constant
  from the $t \bar{t}$ Production Cross Section in $pp$ Collisions at
  $\sqrt{s}$ = 7 TeV}}.
\bjtitle{Phys. Lett. B}
\bvolume{728},
\bfpage{496}--\blpage{517}
(\byear{2014})
{\href{https://arxiv.org/abs/1307.1907}{{arXiv:1307.1907}}}
{[hep-ex]}.
\doiurl{10.1016/j.physletb.2013.12.009}.
\bcomment{[Erratum: Phys.Lett.B 738, 526--528 (2014)]}
\end{barticle}
\endbibitem

\bibitem{Klijnsma:2017eqp}
\begin{barticle}
\bauthor{\bsnm{Klijnsma}, \binits{T.}},
\bauthor{\bsnm{Bethke}, \binits{S.}},
\bauthor{\bsnm{Dissertori}, \binits{G.}},
\bauthor{\bsnm{Salam}, \binits{G.P.}}:
\batitle{{Determination of the strong coupling constant $\alpha_s(m_Z)$ from
  measurements of the total cross section for top-antitop quark production}}.
\bjtitle{Eur. Phys. J. C}
\bvolume{77}(\bissue{11}),
\bfpage{778}
(\byear{2017})
{\href{https://arxiv.org/abs/1708.07495}{{arXiv:1708.07495}}}
{[hep-ph]}.
\doiurl{10.1140/epjc/s10052-017-5340-5}
\end{barticle}
\endbibitem

\bibitem{ATLAS:2019hau}
\begin{barticle}
\bauthor{\bsnm{Aad}, \binits{G.}}, \betal:
\batitle{{Measurement of the $t\bar{t}$ production cross-section and lepton
  differential distributions in $e\mu $ dilepton events from $pp$ collisions at
  $\sqrt{s}=13\,\text {TeV}$ with the ATLAS detector}}.
\bjtitle{Eur. Phys. J. C}
\bvolume{80}(\bissue{6}),
\bfpage{528}
(\byear{2020})
{\href{https://arxiv.org/abs/1910.08819}{{arXiv:1910.08819}}}
{[hep-ex]}.
\doiurl{10.1140/epjc/s10052-020-7907-9}
\end{barticle}
\endbibitem

\bibitem{CMS:2016qqr}
\begin{barticle}
\bauthor{\bsnm{Khachatryan}, \binits{V.}}, \betal:
\batitle{{Measurement of the differential cross section and charge asymmetry
  for inclusive $\mathrm {p}\mathrm {p}\rightarrow \mathrm {W}^{\pm }+X$
  production at ${\sqrt{s}} = 8$ TeV}}.
\bjtitle{Eur. Phys. J. C}
\bvolume{76}(\bissue{8}),
\bfpage{469}
(\byear{2016})
{\href{https://arxiv.org/abs/1603.01803}{{arXiv:1603.01803}}}
{[hep-ex]}.
\doiurl{10.1140/epjc/s10052-016-4293-4}
\end{barticle}
\endbibitem

\bibitem{BCDMS}
\begin{barticle}
\bauthor{\bsnm{Benvenuti}, \binits{A.C.}}, \betal:
\batitle{{A High Statistics Measurement of the Proton Structure Functions F(2)
  (x, Q**2) and R from Deep Inelastic Muon Scattering at High Q**2}}.
\bjtitle{Phys. Lett.}
\bvolume{B223},
\bfpage{485}--\blpage{489}
(\byear{1989}).
\doiurl{10.1016/0370-2693(89)91637-7}
\end{barticle}
\endbibitem

\bibitem{ATLAS:2017rue}
\begin{barticle}
\bauthor{\bsnm{Aaboud}, \binits{M.}}, \betal:
\batitle{{Measurement of the Drell-Yan triple-differential cross section in
  $pp$ collisions at $\sqrt{s} = 8$ TeV}}.
\bjtitle{JHEP}
\bvolume{12},
\bfpage{059}
(\byear{2017})
{\href{https://arxiv.org/abs/1710.05167}{{arXiv:1710.05167}}}
{[hep-ex]}.
\doiurl{10.1007/JHEP12(2017)059}
\end{barticle}
\endbibitem

\bibitem{Czakon:2013tha}
\begin{barticle}
\bauthor{\bsnm{Czakon}, \binits{M.}},
\bauthor{\bsnm{Mangano}, \binits{M.L.}},
\bauthor{\bsnm{Mitov}, \binits{A.}},
\bauthor{\bsnm{Rojo}, \binits{J.}}:
\batitle{{Constraints on the gluon PDF from top quark pair production at hadron
  colliders}}.
\bjtitle{JHEP}
\bvolume{07},
\bfpage{167}
(\byear{2013})
{\href{https://arxiv.org/abs/1303.7215}{{arXiv:1303.7215}}}
{[hep-ph]}.
\doiurl{10.1007/JHEP07(2013)167}
\end{barticle}
\endbibitem

\bibitem{Workman:2022ynf}
\begin{barticle}
\bauthor{\bsnm{Workman}, \binits{R.L.}},
\bauthor{\bsnm{Others}}:
\batitle{{Review of Particle Physics}}.
\bjtitle{PTEP}
\bvolume{2022},
\bfpage{083}--\blpage{01}
(\byear{2022}).
\doiurl{10.1093/ptep/ptac097}
\end{barticle}
\endbibitem

\bibitem{Li:2013mia}
\begin{barticle}
\bauthor{\bsnm{Li}, \binits{H.T.}},
\bauthor{\bsnm{Li}, \binits{C.S.}},
\bauthor{\bsnm{Shao}, \binits{D.Y.}},
\bauthor{\bsnm{Yang}, \binits{L.L.}},
\bauthor{\bsnm{Zhu}, \binits{H.X.}}:
\batitle{{Top quark pair production at small transverse momentum in hadronic
  collisions}}.
\bjtitle{Phys. Rev. D}
\bvolume{88},
\bfpage{074004}
(\byear{2013})
{\href{https://arxiv.org/abs/1307.2464}{{arXiv:1307.2464}}}
{[hep-ph]}.
\doiurl{10.1103/PhysRevD.88.074004}
\end{barticle}
\endbibitem

\bibitem{Catani:2018mei}
\begin{barticle}
\bauthor{\bsnm{Catani}, \binits{S.}},
\bauthor{\bsnm{Grazzini}, \binits{M.}},
\bauthor{\bsnm{Sargsyan}, \binits{H.}}:
\batitle{{Transverse-momentum resummation for top-quark pair production at the
  LHC}}.
\bjtitle{JHEP}
\bvolume{11},
\bfpage{061}
(\byear{2018})
{\href{https://arxiv.org/abs/1806.01601}{{arXiv:1806.01601}}}
{[hep-ph]}.
\doiurl{10.1007/JHEP11(2018)061}
\end{barticle}
\endbibitem

\bibitem{Alioli:2021ggd}
\begin{barticle}
\bauthor{\bsnm{Alioli}, \binits{S.}},
\bauthor{\bsnm{Broggio}, \binits{A.}},
\bauthor{\bsnm{Lim}, \binits{M.A.}}:
\batitle{{Zero-jettiness resummation for top-quark pair production at the
  LHC}}.
\bjtitle{JHEP}
\bvolume{01},
\bfpage{066}
(\byear{2022})
{\href{https://arxiv.org/abs/2111.03632}{{arXiv:2111.03632}}}
{[hep-ph]}.
\doiurl{10.1007/JHEP01(2022)066}
\end{barticle}
\endbibitem

\bibitem{Ju:2022wia}
\begin{barticle}
\bauthor{\bsnm{Ju}, \binits{W.-L.}},
\bauthor{\bsnm{Sch\"onherr}, \binits{M.}}:
\batitle{{Projected transverse momentum resummation in top-antitop pair
  production at LHC}}.
\bjtitle{JHEP}
\bvolume{02},
\bfpage{075}
(\byear{2023})
{\href{https://arxiv.org/abs/2210.09272}{{arXiv:2210.09272}}}
{[hep-ph]}.
\doiurl{10.1007/JHEP02(2023)075}
\end{barticle}
\endbibitem

\end{thebibliography}


\end{document}